\documentclass[aps,pre,twocolumn,showpacs,epsfig,epsf,amssymb,floatfix]{revtex4-1}
\usepackage{xcolor}\usepackage{graphicx}
\usepackage{amssymb}
\usepackage{amsmath}
\usepackage{subfigure}
\usepackage{booktabs}
\usepackage{float}
\usepackage{longtable}
\begin{document}
\title{Transient helix formation in  charged semiflexible polymers without confinement effects.}
\author{Debarshi Mitra$^{1}$, Apratim Chatterji$^{1,2}$}
\email{apratim@iiserpune.ac.in}
\affiliation{
$^1$ Dept. of Physics, IISER-Pune, Dr. Homi Bhaba Road,  Pune-411008, India.\\
$^2$ Center for Energy Science, IISER-Pune,  Dr. Homi Bhaba Road,  Pune-411008, India.
}

\date{\today}
\begin{abstract}
Switching on generic interactions e.g. the Coulomb potential or other long ranged
spherically symmetric repulsive  interactions  between monomers of bead-spring
model of a semi-flexible polymer induce instabilities in a semiflexible polymer
chain to form  transient helical structures. 
{ 
Our proposed mechanism
could explain the spontaneous emergence of helical order in stiff (bio-) polymers 
as a chain gets charged from a neutral state. But since the obtained helical structures 
 dissolve away with time, hydrogen bonding (or  other additional mechanisms),
would be required to form stabilized helical structures as observed 
in nature (such as in biological macro-molecules)}. The emergence of the
helix  is independent of the molecular details of the monomer constituent. 
The key factors  which control the emergence of the helical structure is the
persistence length and  the charge density. We have avoided using torsional
potentials to obtain the transient helical structures. 
Moreover, we can drive the semiflexible polymer to form helices
in a recurring manner by periodically increasing and decreasing the effective
charge of the monomers. 
{ If the two polymer ends are tethered to two surfaces 
separated by a distance equal to the contour length of the polymeric chain, which could be in the range $10 nm$-$\mu$, 
the life time 
of the helical structures formed is increased.}
\end{abstract}

\keywords{Self-assembly and self-organization, polymer physics, emergent helices}

\maketitle
\section{Introduction}

Creating emergent structures through intelligent engineering  of physical
interactions between macro-molecules is a versatile method to self-assemble or 
self-organize structures with a target morphology.  A particular macro-molecular 
morphology of great interest across disciplines is the helix, as it is a recurring
motif across chemistry, biology\cite{Banavar2004,Gerbode2012,Forterre2011} and
physics \cite{Maritan2000,Pokroy2009,Sabeur2008,Li2018,Vega2019,Snir2005}. Forging helical 
structures at the $nm$ to $10 \mu$ length scales remain challenging, though 
helical springs are ubiquitous  in NEMS/MEMS devices\cite{Huang2015,Gao2006}
,piezoelectric devices\cite{Kong2003}
and  helical micro-swimmers\cite{Tottori2012,Ghosh2009,Zhang2009,Zhang2010} are used for
micro-rheology. These helices are produced  primarily by various ``bottom up approaches'', 
e.g. vapour deposition which is dependent on the detailed interactions of the 
constituent atoms/ molecules, or alternatively using  helical
templates\cite{Wang2013,Wang2008,Zhang2002,Zhang2003,Liu2011,Robbie1999,Rapaport2002}. 
Helices can also emerge due to suitable confinement effects \cite{Chaudhuri2012,Sanwaria2015}.
It would be of   interest to devise  alternate strategies to obtain spontaneously formed 
helical architectures at $nm-\mu$ length scales using  physical  forces by approaches 
which remain independent of chemical details of the monomer constituent.

There have been previous reports of extremely short lived helix
formation in polymers in bad solvents undergoing collapse due to
hydrophobic forces \cite{Sabeur2008} which act at $nm$ length scales.
Others have observed helices on optimally packing tubular filaments
at  particular  ratios of pitch and radius \cite{Maritan2000}. But 
{ in a more detailed paper,} the authors comment that compaction of a chain of spheres gives 
very different results from compaction of a tube \cite{Banavar2004}. 
This is because the tube can be considered as a compact object made 
up of discs which has very different symmetry properties in terms 
of interaction potentials compared to those acting between spherical
beads,  say, of a polymeric chain. Another study shows that the
ground state of a self attracting  chain shows a variety of 
structural motifs including the helix, depending on  the nature of
the stiffness present in the chain (energetic or entropic) 
\cite{krbi2019}.  Our study reports  the self-emergence of free
standing helical structures using the most generic of repulsive 
potentials such as Coulomb repulsion  which could have  influence 
in understanding emergence of such structures at  nm-$\mu$ 
length scales, in a variety of situations within the living cell or outside.

Here we show emergent  structures with transient helical order in a
free standing (unconfined) bead spring model of a semiflexible polymeric
chain using generic interactions. Our computer simulations
 show that  helical  structures can be obtained by
inducing instabilities with either Coulomb interactions  or
other long ranged power law  repulsive interactions between
the monomers  if we start out from an initial
configuration where the uncharged polymer chain is straight {
for a polymer chain whose persistence length is lesser than the contour length}. 
At time $t=0$ if a neutral polymer becomes charged, the chain adopts a helical
configuration before the helical structure dissolves to adopt a stretched
linear configuration at long times due to long ranged repulsion. 
{ We also show that a stiff polymer chain in thermal 
equilibrium with its 
bath can also result in  a helical conformation if repulsive Coulomb 
interactions between the monomers is switched on}. 
{ A helical structure may 
also be obtained if a semi-flexible polymer  chain,  with persistence length 
$\ell_p < N_c$ where $N_c:$ the contour length) is pulled at both 
ends by a constant force, and released just as the repulsive Coulomb interaction 
is switched on between the monomers. Experimentally this may be accomplished by 
pulling the polymer chain with an AFM tip} {
\cite{Haupt2002,Oesterhelt1999}} and the charges may be induced by 
changing the  pH of the solution, 
{ \cite{Kirwan2004,Katchalsky1957,Kocak2017,Suresh2019,Kumaraswamy2020}}.  

Note that in all of the above we obtain transient helices without the use of
torsion inducing  potentials or hydrogen-bond mimicking  potentials acting between
monomers. 
{ In this paper}, we also show how thermal fluctuations play
an important  role in the the formation of helical structures. In addition, 
we induce time dependent potentials where the charge  of a semi-flexible 
polymer varies with time (say as pH changes with time){ \cite{Orbn2015,Dodd2017}} . As a consequence, 
helices are formed periodically in phase with
the driving.

The manuscript is organized as follows. The following
section  describes the model of a semi-flexible polymer
dressed with additional long range interactions which leads
to helix formation. The additional interactions have the
form $\sim 1/r$,  or $\sim 1/r^3$.  This implies that there is 
no screening of Coulomb charges, when we describe helix formation 
starting out from a  straight line initial condition or of a stiff 
polymer in thermal equilibrium or from a stretched condition due
to force applied  to the end monomers. Next we discuss the 
mechanism of  helix formation starting out from a  straight line
initial condition (for simplicity) with $1/r$ (Case A) and 
$1/r^3$ (Case B). At the end we discuss the the range of values 
of semi-flexibility energies/spring constants/strength of Coulomb
forces for which we obtain helices. We do this by plotting a suitable 
state-diagram. We finally conclude with Discussions and future outlook.

\section{Model}
We use  the bead spring model of a polymer for our simulations. 
The  model  polymer could be a real polymer, or it can be string 
of colloids stitched together to form a semi-flexible polymeric chain 
as described in  \cite{Biswas2017,Kumaraswamy2016}. Thereby, the 
monomer size and the number of beads $N$ in the chain determine the 
length scale of helical configurations formed.  The unit of length 
in our study is $a$, where $a=1$  is the equilibrium length of the
harmonic-springs { between two adjacent monomers} with energy  $u_H = \kappa (r - a)^2$ between
adjacent monomers; $r$ is the distance between the monomers. The 
 spring constant $\kappa$ is $20  k_BT/a^2$ for Case A, which has 
 repulsive Coulomb interactions $u_c= \epsilon_c (a/r)$  
 acting between  all monomer pairs of the chain.  The  parameter
 $\epsilon_c = 87.27k_B T$ is the measure
of the Coulomb energy when a pair of charges are at a distance $a$ 
from each other. Case B  uses $\kappa=10 k_BT/a^2$,  along with the
additional interaction $u_d$  between  all the monomers
of the   chain.  The form of the potential $u_d$ is
$u_d= \epsilon_d (a/r)^3$  with $\epsilon_d = 107.70 k_BT$ 
with cutoff at $r_c = 4a$. Diameter of each  monomer is 
$\sigma=0.727a$, and excluded volume (EV) of  monomers are modeled 
by the WCA (Weeks Chandler Anderson) potential.{
This choice corresponds to the good solvent condition.}

The polymeric chain is semi-flexible; the corresponding
bending energy $u_b$ is $u_b=\epsilon_b cos(\theta)$,  where
$\theta$ is the angle between vectors ($-\mathbf{r}_i,
\mathbf{r}_{i+1}$). The vector $\mathbf{r}_{i}$ is the vector 
joining  monomer $i-1$  to its neighbouring monomer $i$
along the chain contour.  The thermal energy $k_BT=1$ sets the
energy unit. { We performed  Brownian dynamics
simulations where the friction constant is $\mathbf{\zeta}$, 
and the unit of time $\tau$ is set by $\tau=a^2 \zeta/k_BT$, the time taken  for a isolated monomer
particle to diffuse a distance  of $a$..
If we set $\zeta=1$ such that $\tau=1$, since $k_bT$ and $a$ are 
already chosen to be one, the over-damped stochastic Brownian 
dynamics equation is integrated with time step $dt =0.0001 \tau$}. 
  
Unless clarified otherwise, we mostly observe the polymer 
dynamics  by   starting out from the same straight 
line initial  condition  for the above mentioned cases (A)
and (B): a linear  polymer  chain of $49$ monomers is 
placed along the y axis  with  adjacent  monomers at a 
distance of $a$ from each other. The fluctuation dissipation 
theorem determines the magnitude of the  random force  on 
each particle for all cases.  For studies with cases (A) and 
(B),  we choose $\epsilon_b = 10 k_BT$ (corresponding 
to the persistence length $\ell_p = 11a$, 
{ as calculated by simulations}) 
and  $80 k_BT$, respectively. A large difference in the values of
$\epsilon_b$ was  chosen to demonstrate that helix formation 
is robust for a range of  parameter values. We use box-size 
$\gg 50a$ for a chain with $49$  beads, such that periodic 
boundary conditions are never invoked. Hence  we do not use 
Ewald technique to  calculate Coulomb interactions as self 
interactions with periodic images of the monomers are 
irrelevant.  Moreover, no counterions were considered for 
our simulations, so the charges are not screened.

Subsequently, { we study transient helix formation of 
chains in thermal 
equilibrium, {\em viz.},}  we establish that a semiflexible polymer with $60$ 
monomers in  the chain and with  persistence length $\ell_p$ greater than 
the contour length { $N_c = 60a$} and in thermal equilibrium with the bath,  
develops a local helical order once the repulsive Coulomb interactions
($\epsilon_c=87.27k_BT$) between the monomers is switched on. The 
spring constant $\kappa=200 k_BT/a^2$. Furthermore, if a semi-flexible 
polymer, with 60 monomers in the chain but with $\ell_p <60a$, is 
stretched by applying a constant force at both ends by a constant 
force $20k_BT/a$ and then released, and  simultaneously the repulsive 
Coulomb interaction ($\epsilon_c=87.27k_BT$) is switched on between 
the monomers, then the polymer again develops a transient helical order. 

From the experimental perspective, it would be more instructive to 
specify the $\ell_p$ of a polymer rather than  specify the the 
simulation parameter $\epsilon_b$, which we { tune} 
to fix $\ell_p$. To that end, we  calculate the
relation between $\epsilon_b$ and $\ell_p$ for small  angular deviations of
bond-angles from angle $\pi$. The calculation details 
are given in the appendix. The relation between angle $\alpha$  (as shown 
in the Fig.\ref{fig14} of appendix) and  $\epsilon_b$ is given by,
\begin{equation}
 (\epsilon_b'-1)/\epsilon_b'=\cos \alpha
\end{equation}
where $\epsilon_b'={\epsilon_b/k_BT}$
and $\alpha=\pi -\theta$. From polymer physics \cite{Rubinstein},
we know that for  WLC (worm like chain) model, for the small 
angles  of bends, the persistence length $\ell_p$ is given by
$\ell_p = 2a/\alpha^2$. Thereby,
\begin{equation}
    \ell_p \approx a \epsilon_b/k_BT
\end{equation}
Thus $\ell_p$ increases linearly with $\epsilon_b$. As an example, 
a polymer with bending energy $\epsilon_b=10 k_BT$ will have 
persistence length $\ell_p \approx 10a$ as per the above equation. This 
matches with the  earlier mentioned value of $\ell_p=11a$, where we 
explicitly calculated the $\ell_p=11a$ from the decay of the  correlation
function of the end-to-end vector for a semi-flexible polymer (with $u_c$ 
kept fixed at 0). At higher values  of $\epsilon_b$, the Eqn.2 will be
more accurate.

\section{Results and Analysis:}
\begin{figure}[!hbt]
\includegraphics[width=0.48\columnwidth]{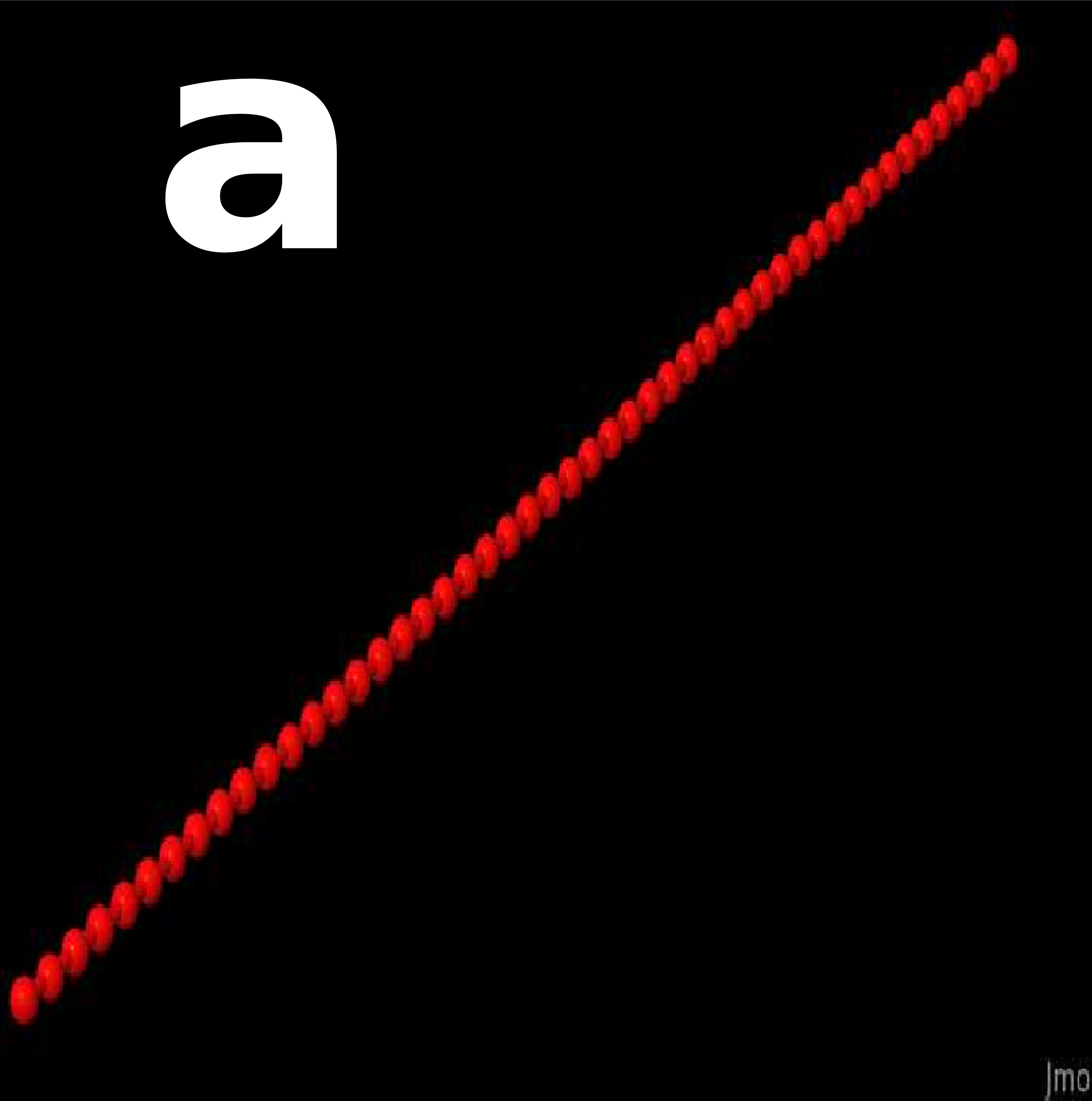}
\hskip0.1cm
\includegraphics[width=0.48\columnwidth]{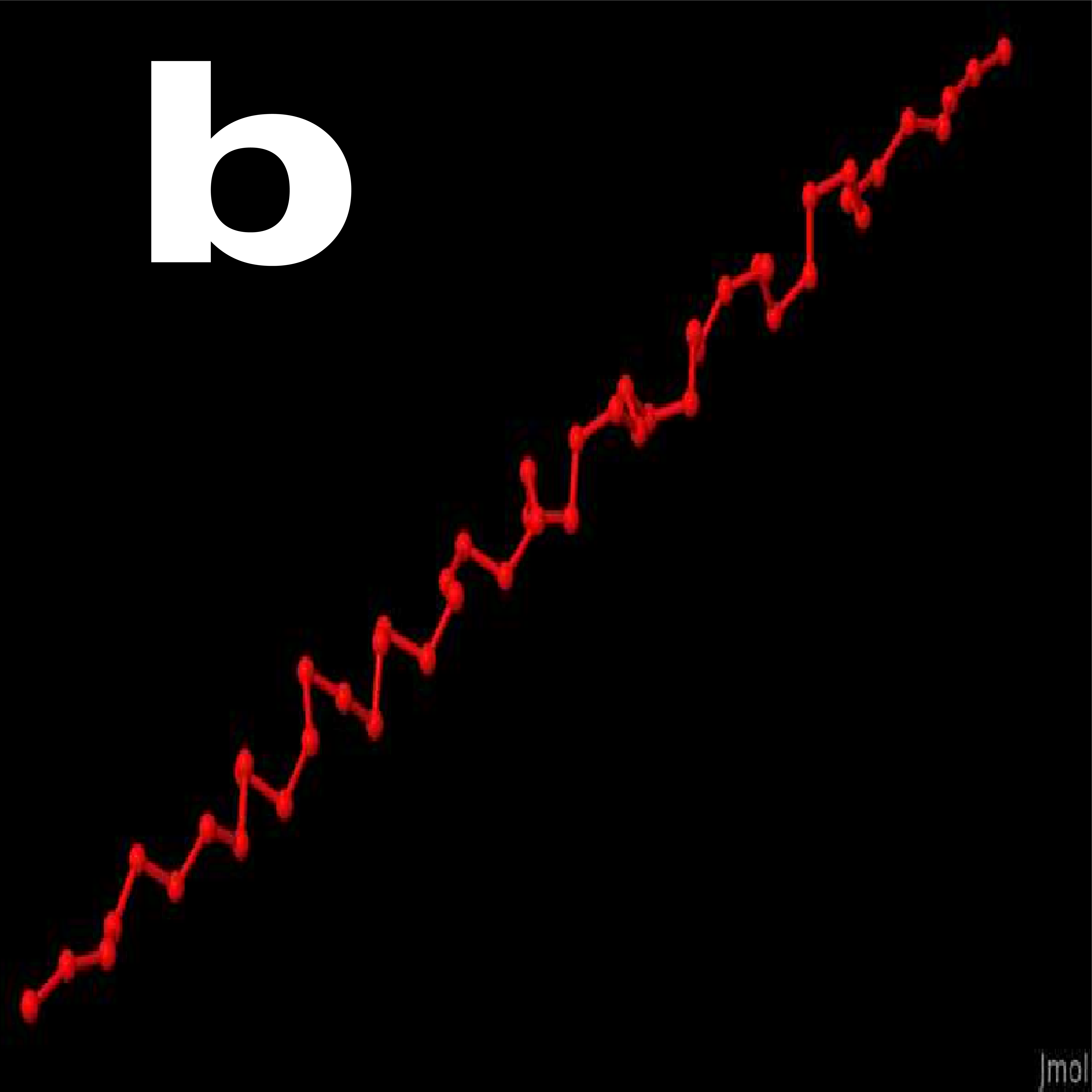}\\
\vskip0.5cm
\includegraphics[width=0.48\columnwidth]{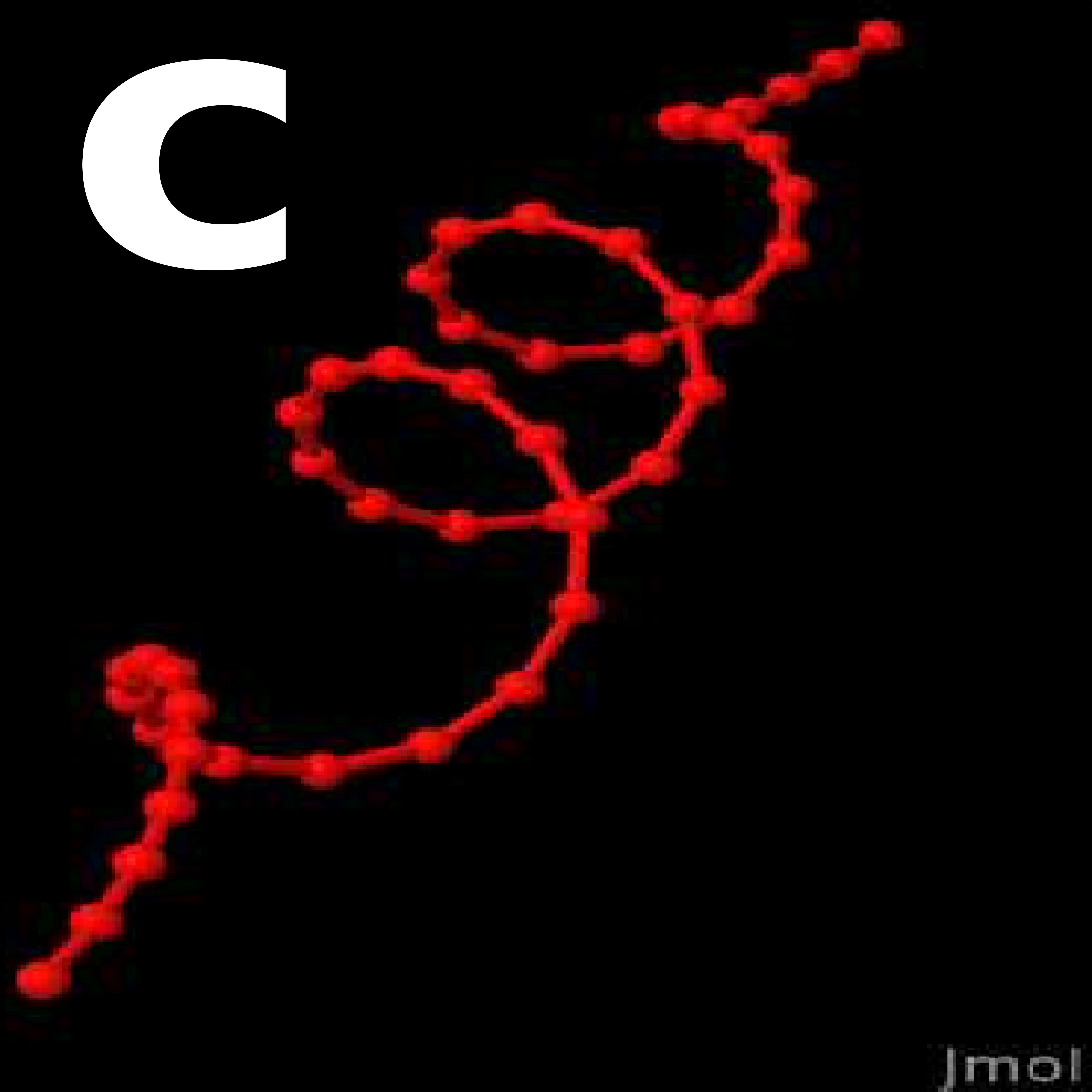}
\hskip0.1cm
\includegraphics[width=0.48\columnwidth]{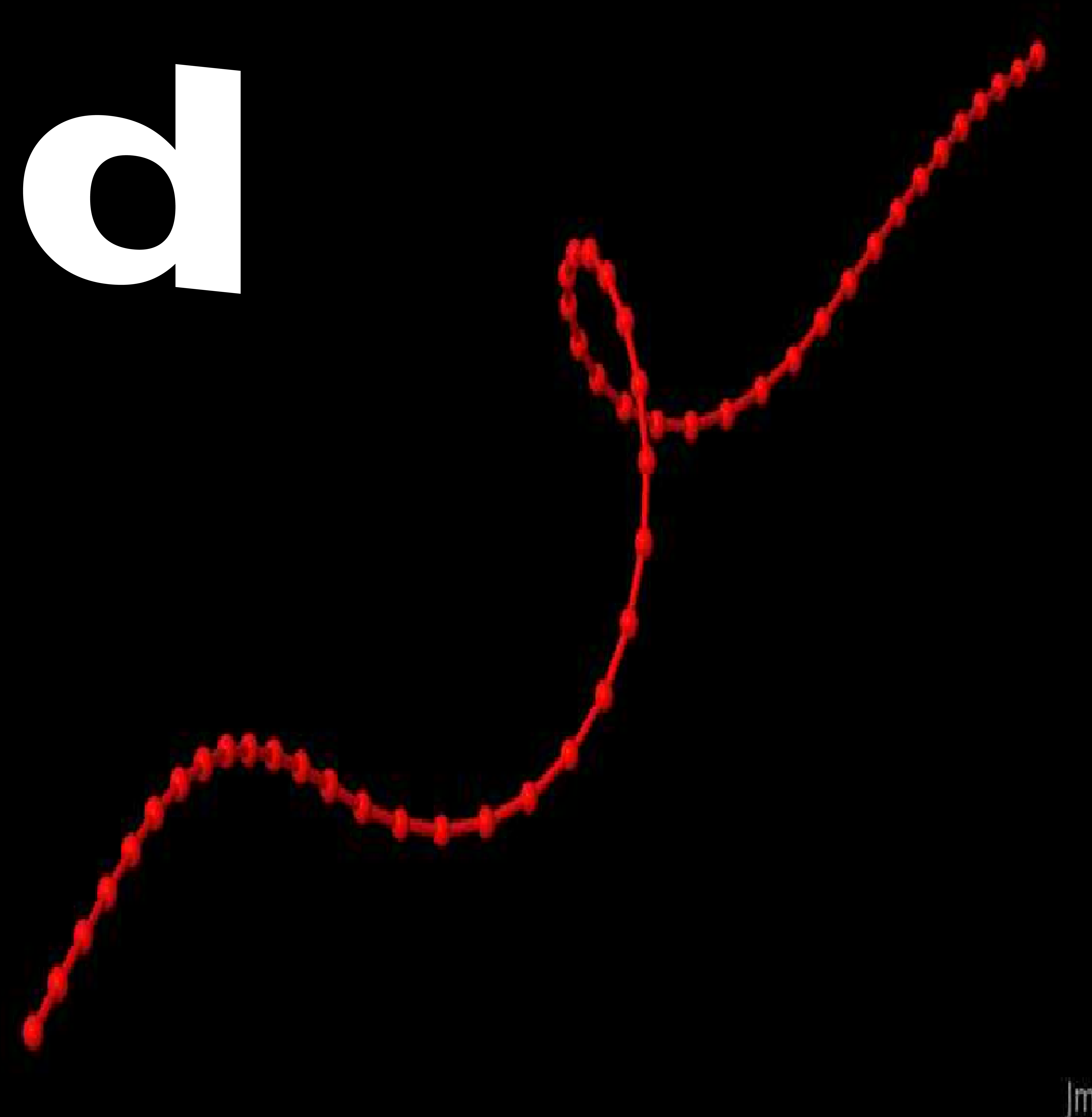}
\vskip0.5cm
\caption{\label{fig1}
Figures 1(a-d) shows various stages of the helical instability for a 
semi-flexible polymeric chain starting from the straight initial
configuration with potential $u_d$: case B. The snapshots are
(a) for the { straight line} initial configuration at time $t=0$ 
with 49 monomers (b) the  configuration at 
time $t=3.3 \times 10^{-2} \tau$ { ($\mathbf {H2=0.23, H4=0.006}$)}
(c) the configuration at a subsequent time
$t=\tau$,  when the helix is formed {$({\mathbf H2=0.81,H4=0.43})$}
(d) configuration showing the unwinding of the helix  at time $t=5\tau$
{$(\mathbf{H2=0.65,H4=0.29})$}. The corresponding  snapshots with potential 
${u_c}$ (Case A) are in the supplementary.}
\end{figure}

In Figure \ref{fig1} we show representative snapshots from various stages 
of transient  helix formation for  a polymer with interaction energies
corresponding to case-B starting out from a straight initial condition.
As the bead-spring model of polymer chain starts out from  the straight line 
initial configuration (refer Fig.\ref{fig1}a), the thermal forces randomly
displace the monomers from their initial  positions. Furthermore, strong
repulsive  forces arising from $u_d$  act  along the line joining  
the centers of monomers  make the monomers move  away from each other,
accentuating the angle between adjacent bonds and consequently
the polymer forms a locally kinked structure as shown in Fig.\ref{fig1}b 
which is penalized by the bending energy term. Thereby, the helical 
conformation  of the polymer emerges at sections of the chain at a
time $t \sim \tau$   to  locally relax the the high bending energy costs 
due to kinks as seen in Fig.\ref{fig1}c.  But it dissolves  away at 
times  $t>>\tau$ (refer Fig.\ref{fig1}d). The unit  of time  of the problem 
is chosen as  $\tau = (\zeta a^2/k_BT)$, the time taken  for a isolated monomer
particle to diffuse a distance  of $a$.

Random fluctuations due to $k_BT$  displace the monomers just 
after time $t=0$, which in turn leads to the  development of the
helical order, and thereby $k_BT$ plays a crucial role  though 
$\epsilon_c, \epsilon_d$  and $\kappa, \epsilon_b$ are all $\gg k_BT$. 
A  perfectly straight polymer configuration at $T=0$ stretches out 
but never gets to form helices as all the forces  between monomers 
act along the line joining the centers.  Movies S1, S2
in the Supplementary section helps the reader to visualize 
the instability  which results in helix formation for
Case-A $\&$ Case-B, respectively. Movie S3  is for Case B 
with potential $u_d$ with $k_BT=0$,  and as a consequence
the polymer does not form transient helical structures.

\begin{figure}[!hbt]
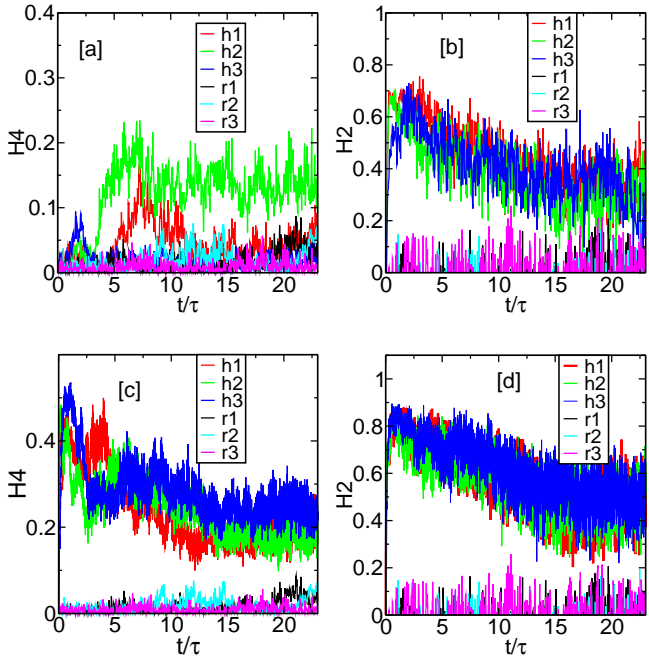

\includegraphics[width=0.48\columnwidth]{files_images/h4_long_run_coulomb.eps}
\hskip0.1cm
\includegraphics[width=0.48\columnwidth]{files_images/h2_long_run_coulomb.eps}
\vskip0.5cm
\includegraphics[width=0.48\columnwidth]{files_images/h4_long_run_dipole.eps}
\hskip0.1cm
\includegraphics[width=0.48\columnwidth]{files_images/h2_long_run_dipole.eps}
\vskip0.5cm
\caption{\label{fig2}
 Subfigures (a),(b) shows $H4$ and $H2$ versus time  $t$ of a 
 semi-flexible polymer chain of $49$ monomers for three independent runs denoted by 
 $h1,h2$ and $h3$ starting out from  a straight  initial configuration of the chain. 
 Potential $u_c$ acts between all monomers pairs. The interaction strengths 
 correspond  to Case A. For comparison,  we also show $H4,
 H2$ values obtained for a semi-flexible polymer chain of $49$ monomers  with  $u_c=0$ starting
 from the same initial condition; these are denoted by $r1, r2$ and $r3$.
 Subplots (c),(d) shows $H4$ and $H2$ versus time  of the semi-flexible 
 polymer chain of $49$ monomers for three independent runs denoted by $h1,h2$ and $h3$, such that potential 
 $u_d$ acts between all monomer pairs;  the interaction energies 
 correspond to case B. The  initial configuration  is a straight chain 
 along y axis. The three independent runs for a chain of same length with $u_d=0$  are denoted by 
 $r1, r2$ and $r3$. Note that $H2$ is equal to average
of the values of$\cos(\phi_i)$; $\phi_i$ are the torsion angles subtended 
along the chain contour. In each figure, data for $H2,H4$ is plotted every
$1000$ iterations, i.e.  every $0.1 \tau$.}
\end{figure}

We quantify the emergence of helicity as a function of time by
calculating and plotting two quantities in Fig.\ref{fig2}(a,b)
and Fig. \ref{fig2}(c,d) for cases A and B, respectively,
viz.,  the global order parameter $H4$ and the local order
parameter $H2$ where,
\begin{equation}
 H4= \frac{1}{N-2} (\sum_{i=2}^{i=N-1} \mathbf{u}_i)^2; 
 H2=\frac{1}{N-3}(\sum_{i=2}^{i=N-2} \mathbf{u}_i.\mathbf{u}_{i+1})
\end{equation}
 where $\mathbf{u}_i$ is the unit vector of $ \mathbf{U}_i =
 \mathbf{r}_i \times  \mathbf{r}_{i+1}$. A compact
 tightly wound perfect helix in the continuum picture with
 infinitesimal $\mathbf{r}_i, \mathbf{r}_{i+1}$
 vectors will have vectors $\mathbf{u}_i$ pointing  along 
 the helix axis, and hence $H4$ will have a value of 
 $ \approx 1$.  However, if one obtains a helical structure 
 where half of the chain is right handed, and
 the rest of it is left-handed, $H4$ will be zero. Hence, we
 need the other parameter $H2$ to  identify local helical order 
\cite{Sabeur2008,Kemp2002}.  A simple semi-flexible polymer chain 
($u_c=0$ \& $u_d=0$) shows $H2, \,H4$ values $\approx 0$ (or negative values of $H2$)
as expected for a chain locally bent due to thermal
fluctuations. But  the polymer chains with additional interactions 
$u_c$ or  $u_d$   lead  to the formation  of transient helices
with distinctly non-zero positive values of $H2, H4$. The time
taken for the helix to form  is $\approx  0.5\tau$. Note that $H2$ is 
equal to $\left< \cos (\phi)  \right>$ where $\phi$ denotes the  torsion angle, 
i.e.  the angle  between the planes formed by adjacent pair of the monomer-triplets 
along the  length of the chain; the average is taken over the cosine of the
various torsion angles formed along the length of the chain.

\begin{figure}[!]
\includegraphics[width=0.48\columnwidth]{files_images/end.eps}
\hskip0.1cm
\includegraphics[width=0.48\columnwidth]{files_images/h2_new.eps}
\vskip0.5cm
\includegraphics[width=0.48\columnwidth]{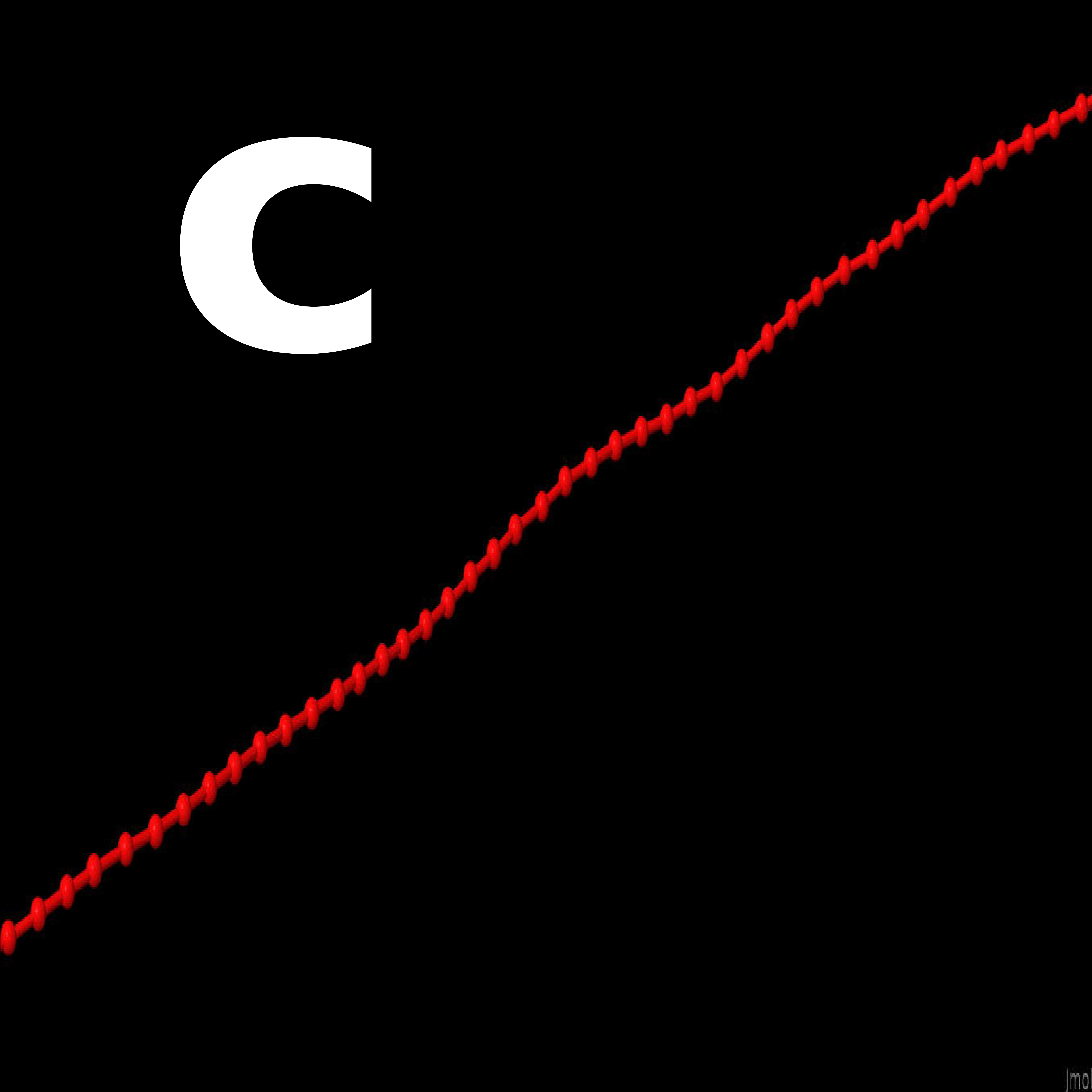}
\includegraphics[width=0.48\columnwidth]{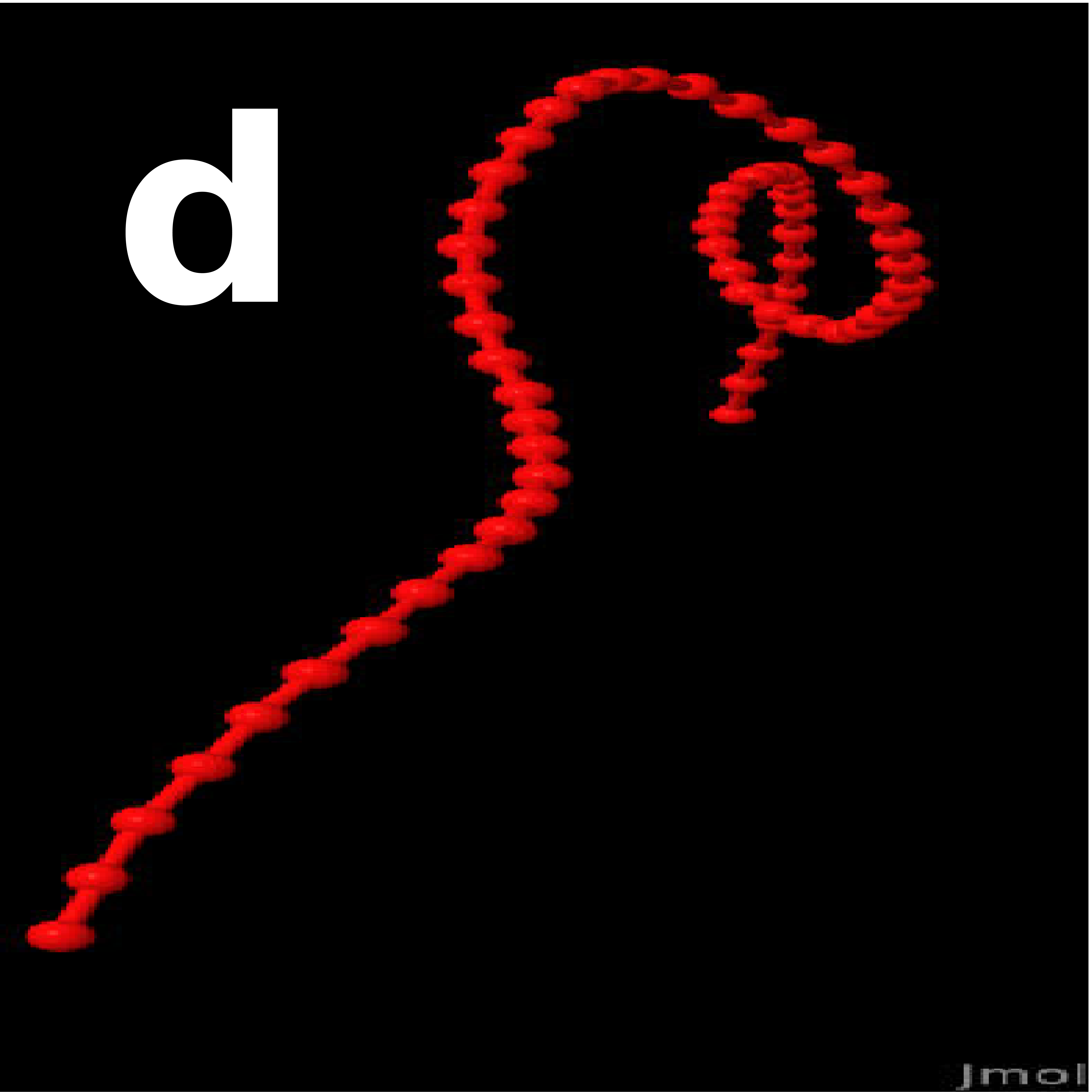}
\caption{\label{fig3}
Subfigure (a) shows the end to end distance $R_{end}$ versus time for 
an uncharged, semiflexible polymer chain of $60$ monomers having 
$\epsilon_b=400 k_BT$  ($\ell_p=400a$) and $\kappa=200 k_BT/a^2$ for 
three independent runs $e1,e2$ and $e3$. Initially the monomers were placed randomly, and  we conclude that $R_{end}$ takes about 
$200 \tau$ to reach its equilibrium value. Subfigure (b) shows $H2$
versus time for the $60$  monomer polymer chain such 
that the repulsive Coulomb interaction ($\epsilon_c=87.27 k_BT$) is 
switched on  at $333.33 \tau$. This led to a increase in the value of 
$H2$ which later decreases  as the helical order dissolves away. The 
data for three independent runs are labelled as $h1, h2$ and $h3$.  
Subfigure (c) shows the snapshot of the polymer configuration just 
before helix formation{$(\mathbf{H2=-0.09,H4=0.008})$}. Subfigure (d) shows the snapshot of the 
conformation of the polymer which has helical order{$(\mathbf{H2= 0.65,H4=0.09})$}  }
\end{figure}

{ We also investigated the emergence  of helices in chains of
 $N=25$ and $N=100$ monomers,  respectively (refer supplementary
 data). A chain of $100$ monomers  has approximately $5$ helical
 segments and thereby has relatively lower values  of $H4$ 
 in some of the independent runs, since  different segments 
 can form helices of opposing handedness.   But the values of  
 $H2$ obtained for $N=25$ or $N=100$ are comparable to
 that obtained for the $N=49$ polymer chain at time $t=2 \tau$.}
 Thus we establish that we indeed get helical conformations 
 in our model semi-flexible polymer as long as we 
 have long-ranged repulsive interactions between the monomers.

\begin{figure}[!]
\includegraphics[width=0.48\columnwidth]{files_images/end_stretch.eps}
\hskip0.1cm
\includegraphics[width=0.48\columnwidth]{files_images/h2_stretch.eps}
\vskip0.5cm
\includegraphics[width=0.48\columnwidth]{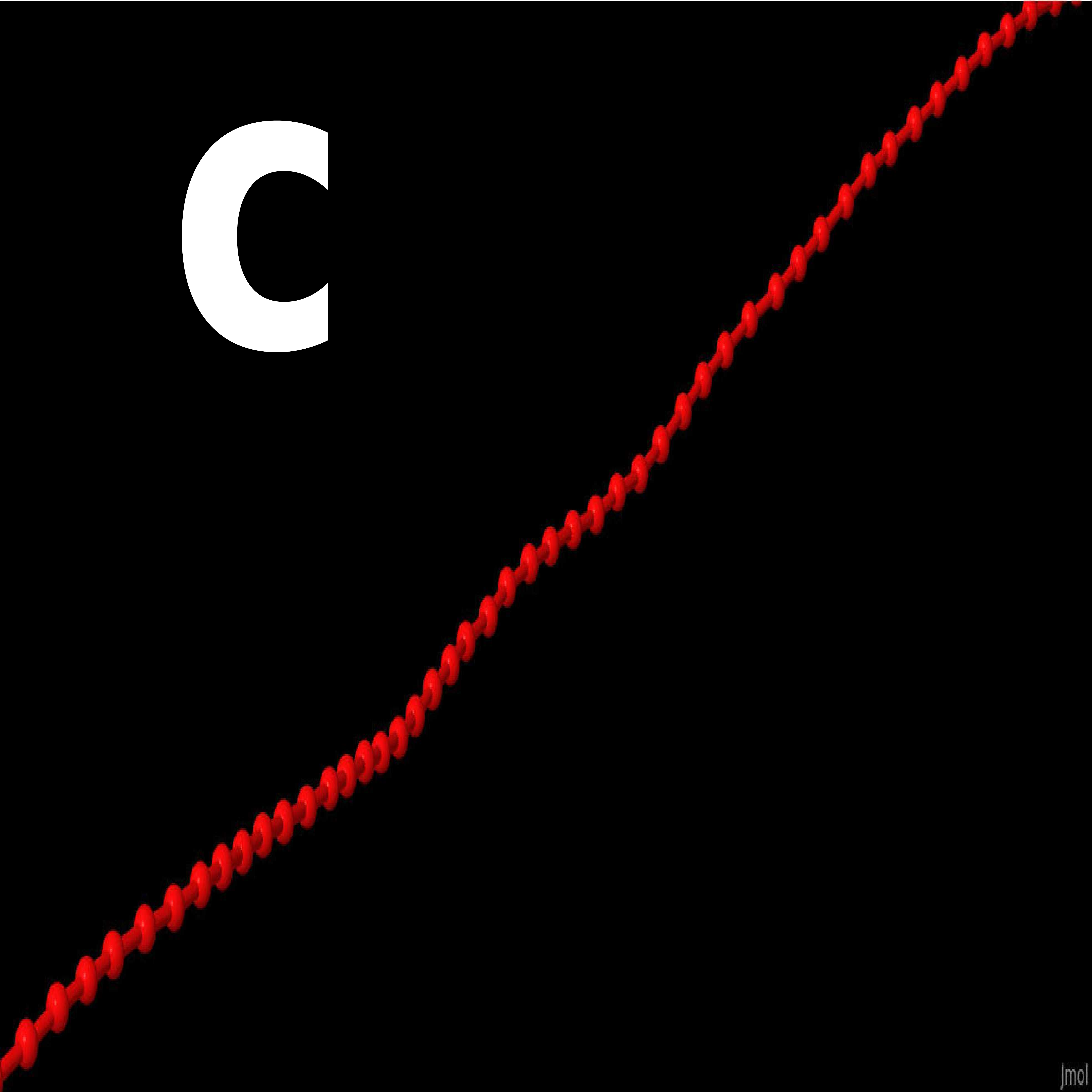}
\includegraphics[width=0.48\columnwidth]{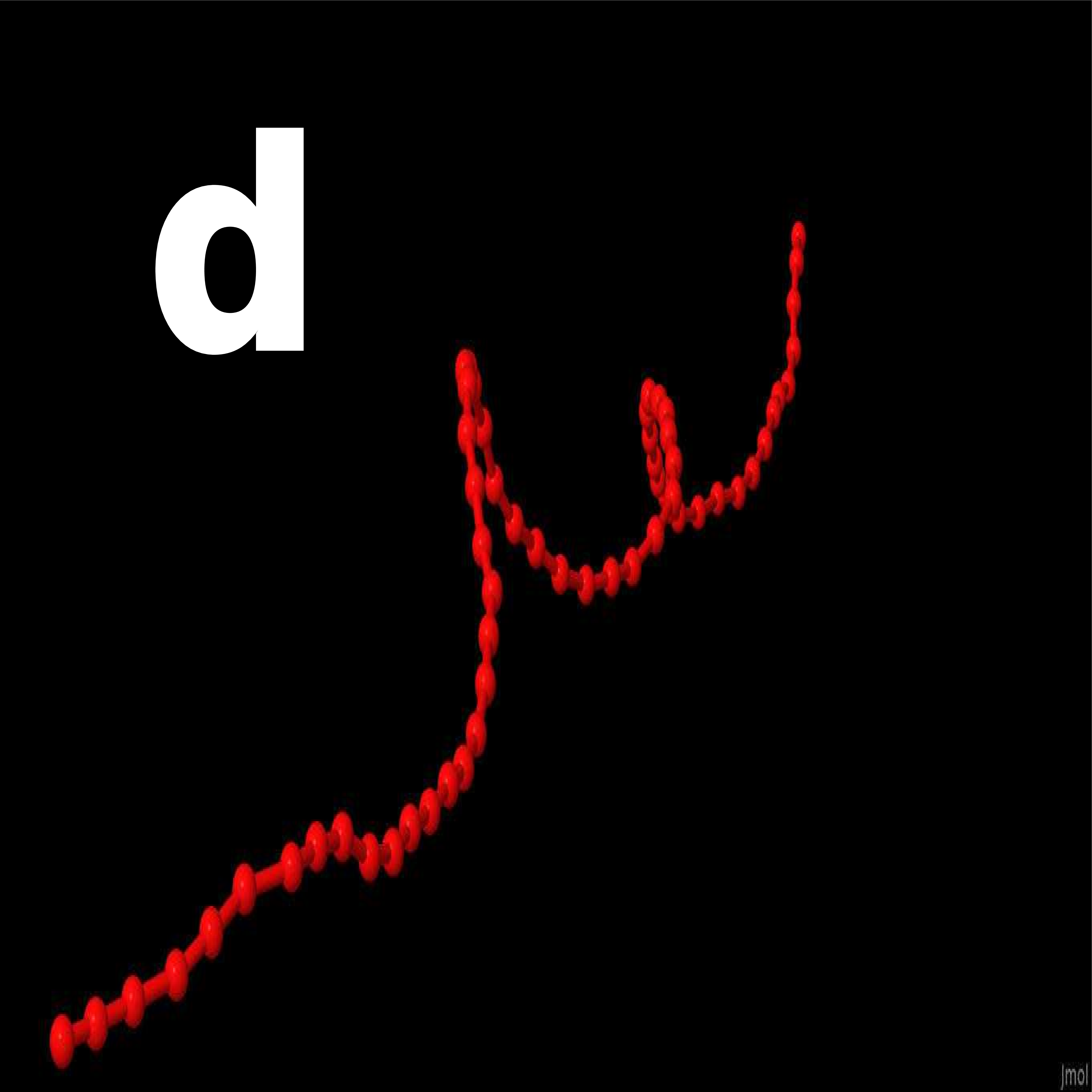}
\vskip0.5cm
\caption{\label{fig4}
Subfigure (a) shows the end to end distance $R_{end}$ versus time $t$
for a polymer chain of $60$ monomers, $\epsilon_b=30 k_BT$ ($\ell_\approx 30a$)
and  $\kappa=200 k_BT/a^2$, where the end monomers are pulled outwards by the 
application of a constant force of $20 k_BT/a$ in opposite directions.
data is presented for 3 independent runs $e1,e2$ and $e3$. Note that the 
Coulomb repulsion ($\epsilon_c=87.27 k_BT$) between monomers was switched
on at $50 \tau$ and the stretching force was set to $0$, simultaneously. 
Subfigure (b) shows $H2$ versus time for  the polymer chains having the 
same parameters values in (a). We once again see transient helix formation 
and its dissolution for three independent runs, $h1, h2$ and $h3$. 
Subfigure (c) shows the snapshot of the polymer configuration just before 
the Coulomb interaction is switched on {$(\mathbf{H2= -0.26,H4=0.012})$}. Subfigure (d) shows the snapshot of 
the helical conformation of the polymer{$(\mathbf{H2= 0.58,H4=0.10})$}}
\end{figure}

{Furthermore, to { establish} that  the helix formation is not 
just a consequence  of the special straight  line initial condition, we  calculate $H2$  to establish the development of helical order in  a semi-flexible polymer
in thermal equilibrium.} The Coulomb potential $u_c$ and corresponding forces between
monomer is switched on after ensuring that the polymer is in equilibrium.
We choose a polymer whose persistence length greater than the contour length, 
place the monomers randomly and allow the polymer to relax and 
reach equilibrium  such that the end to end vector fluctuates about an average 
value. Initially the end-to-end distance $R_{end}$ increases as the bent polymer 
straightens itself. In Fig.\ref{fig3}a we show $R_{end}$ versus time $t$ 
for the semiflexible   polymer chain with $60$ monomers, such that its length 
is $60a$ and $\ell_p= 400a$.  Data is shown for three independent runs, e1, e2 and e3. 
We observe that it takes approximately $200 \tau$  for it to reach the 
equilibrium value. For Fig.\ref{fig3}b we again take the same semiflexible polymer 
with $60$ monomers and switch on the repulsive Coulomb  interaction ($u_c$ with
$\epsilon_c=87.27 k_BT$) between the monomers at $333.33 \tau$. We observe that 
there  is a significant increase in the value of $H2$ immediately after 
$t=333.33 \tau$.  The $H2$ value then gradually decreases, indicating that a 
transient helical structure dissolves away. Fig \ref{fig3}c and Fig\ref{fig3}d 
show the snapshots of the helical conformation just before and after $u_c$ 
was switched on.
 
Suppose we have a polymer chain of length $60a$ with  $\ell_p =30a$ in thermal equilibrium, such that the  persistence length is lower  than the contour length. If we switch on $u_c$,  we do not get any distinct helical order.
However, if we stretch the polymer (say by a AFM-atomic force microscopy tip) 
and then switch on $u_c$, we again see emergence of a transient helical order.
A semi-flexible chain can be stretched by applying a suitable value of
a constant force ($\vec{F}_{\pm} = \pm (20 k_BT/a) \, \hat{y}$) at both ends such 
that its end to end distance $R_{end}$ 
becomes $\approx 60a$. We then allow the chain to explore different 
equilibrium conformations in the presence of the fixed stretching force
acting on the end  monomers. 

Then the  tension is released by switching off the force applied to the end 
monomers and simultaneously  the repulsive Coulomb 
interaction ($u_c$ with $\epsilon_c=87.27 k_BT$) is switched on between the monomers. 
In such a {\em in-silico} experiment, we do see the emergence of helical 
order by  the measurement of $H2$.  The relevant data is shown in 
Fig.\ref{fig4}.  In Fig.\ref{fig4}a we show the evolution of  
$R_{end}$ under the application of 
equal and opposite forces acting on the end monomers of the chain.
The mean $R_{end}$ reaches a mean value greater than the contour length in
three independent runs within time $20 \tau$. At $50 \tau$ there is an increase 
in the end to end distance because at this point the stretching force is 
released  and the repulsive Coulomb interaction (with $\epsilon_c=87.27 k_BT$) 
is switched on. In Fig.\ref{fig4}b we observe that there is a corresponding
significant increase in  the value of $H2$ which gradually decreases indicating
that a transient helical  structure was formed, which dissolves away.
Fig.\ref{fig4}c and Fig.\ref{fig4}d  show the snapshots of the helical
conformation just before and soon after the 
repulsive  Coulomb interaction was switched on between the monomers.     

But what is the physics of helix formation in the semi-flexible polymer chains 
in the presence of spherically symmetric  repulsive  potentials $u_c$ or $u_d$? 
What role does temperature play?  For the remainder of the manuscript, for
simplicity, we report  the dynamics  of a polymer chain in a thermal bath
starting out  from a straight  initial linear conformation.

\begin{figure}[t]
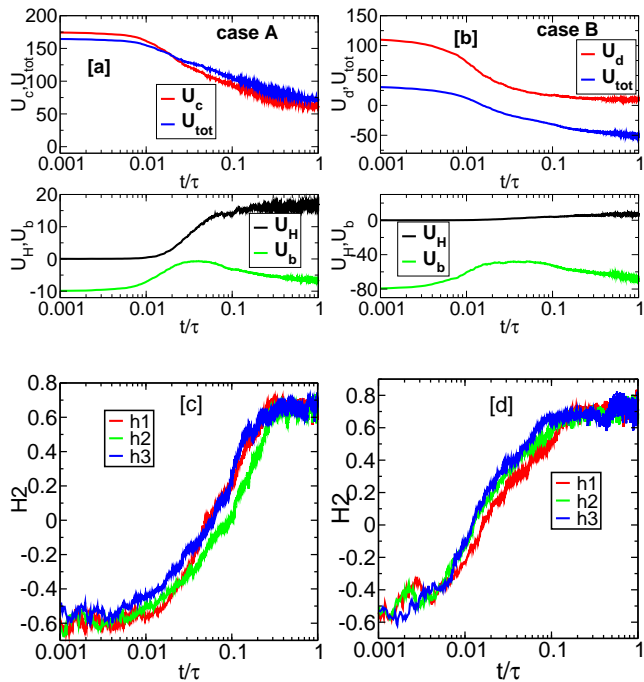

\includegraphics[width=0.48\columnwidth]{files_images/3a.eps}
\includegraphics[width=0.48\columnwidth]{files_images/3b.eps}
\vskip0.5cm
\includegraphics[width=0.48\columnwidth]{files_images/small_t_coulomb_h2.eps}
\includegraphics[width=0.48\columnwidth]{files_images/smallt_dipole_H2.eps}
\vskip0.5cm
\caption{\label{fig5}
 Subplots (a) and (b), corresponding to case A and case B,
 shows energies $U_H$, $U_b$, $U_{tot}$ and  $U_c$ or $U_d$,
 respectively, where $U_{tot}$ denotes the total energy per
 monomer and is the sum of $U_H$, $U_b$ and $U_c$ or $U_d$.
 The x-axis shows $t/\tau$ for relatively short times, i.e.
 $t<\tau$. Subfigures (c) and (d) show $ H2=<cos\phi_i>$ versus
 $t/\tau$ for three independent runs $h1,h2$ and $h3$ with $u_c$
 and $u_d$ acting between the monomers, respectively. 
 The angle $\phi_i$ also denotes the dihedral angle between the  two
 planes formed by the monomers (i,i+1,i+2) and (i+1,i+2,i+3),
 respectively. The index $i$ represents any monomer along the
 chain. The  cosine of the angle $\phi_i$ was averaged along 
 the length of the chain for all possible values of $i$ 
 to yield $ H2=<cos\phi_i>$ at time $t$.}
\end{figure}

  To develop a detailed understanding of  the
 mechanism of helix formation we note that just after time $t=0$
 the thermal kicks   displace the monomers from a straight line
 initial condition.  Thereafter, the magnitude of this random displacements
 gets  accentuated by the repulsive $u_c$ (or $u_d$) acting
 along the line joining the monomer centers, accompanied 
 by an  increase in the distances between monomers. This results
 in the lowering of the Coulomb energy per particle $U_c$ (or
 $U_d$). However, sharp local kinks  get created as is seen in
 Fig.\ref{fig1}b and also results in increase of the contour  length of
the polymer. To release the bending energy due to sharp kinks, the kinked
structure evolves to a structure with local helicity at different segments 
of the chain. We follow the values of the various 
contributions to the total energy $U_{tot}$ as a function of time 
in Fig.\ref{fig5} to understand the development of structure of 
the polymer. Increase in the bond energy per spring, $U_H$ with time
$t/\tau$ indicates  a corresponding stretching of bonds
between adjacent monomers: we have independently checked
that the bonds stretch and do not get compressed. Similarly an increase in 
the value of semi-flexible energy per triplet of monomers, $U_b$, would be
indicative of sharp bends along the contour of the polymer chain.

We now discuss this in more detail. Just after time $t=0$ the chain 
remains nearly straight with the bending energy per each bend nearly 
equal to $-\epsilon_b \cos \theta =-10 k_BT$ since $\cos\theta \approx-1$ 
(case A). However, for time $t/\tau < 0.003$, the $U_H$
increases slowly from value $0$ due to random  shifts in
the monomer positions because of thermal fluctuations.  But this 
increase is not discernible in the plots of the energy contributions 
versus time in Fig.\ref{fig5}, but can be seen in the log-log plot
of energy versus time given in the supplementary section. 
Thereafter, formation of  sharp bends/kinks resulting from 
the motion of monomers due to repulsive Coulomb forces  (or from $u_d$) 
leads to the  rapid increase of both $U_H$ and $U_b$ which is  seen in
\ref{fig5}(a,b) at times $t/\tau > 0.01$. This is  
accompanied by a decrease in the Coulomb energy per 
monomer $U_c$ (and $U_d$), again refer Figs.\ref{fig5}(a,b). 

Following the rapid increase in $U_b$ from time $0.01 < t/\tau < 0.03$, 
there is a sharp increase in forces trying to straighten the chain.
The monomers still move apart from each other due to Coulomb repulsion,
but simultaneously try  to decrease the bending energy  costs by radially 
spreading  out the monomers locally in a manner such that the change in the bending angles along the chain contour becomes
gradual . This dynamics can be deduced by  observing 
the decrease of $U_b$  after it reaches its peak  at time $t/\tau \approx 0.03$.  
As a consequence of the local radial spreading out of the monomers, the chain develops  helicity  along the length of the chain, refer Figs.\ref{fig5}(c,d). 
Note that  the motion of a segment 
would also be constrained  by the motion of adjacent segments
along the chain.  Thus, different segments of the chain could
thus develop clock wise or anti-clock wise helicity since 
the initial  deviations from the straight line configuration 
were in random directions due to thermal fluctuations. 

The evolution of a straight chain into helical structures
can also be followed by looking at the average (along 
the length of the contour) of the average of the cosine of the 
torsion angles along the length of the chain (as given by $H2$), 
as a function of time.
This is plotted in Fig.\ref{fig5}(c) and (d) for cases A and B,
respectively. At time $t=0$, when we have a straight chain,
a plane between monomer triplets is undefined and so is the
normal to the plane. But as soon as the monomers move due to
thermal energy, planes can be defined using the positions of
adjacent monomer-triads  and outward normals $\mathbf{u}_i$ to 
these planes  can point in any  direction but mostly normal to the $y=0$ plane.
At slightly longer times (i.e. when the sharp kinks get
formed), since all values of $\cos(\phi)$ are possible, the 
average of $\cos(\phi)$ along the chain quickly goes to zero
with time $t$ for all the three independent runs. 
However, as the chain develops  helical order beyond time
$t/\tau >0.05$, $\left< \cos (\phi) \right>$ reaches a
values in the range $0.4-0.6$, corresponding to an angle of
around $\sim 60^o$.

\begin{figure}[t]
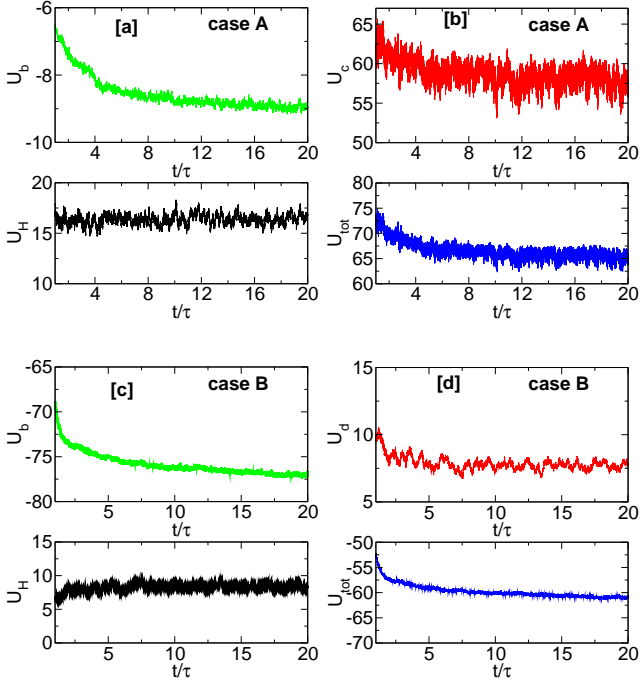

\includegraphics[width=0.48\columnwidth]{files_images/colulomb_en_4.eps}
\includegraphics[width=0.48\columnwidth]{files_images/colulomb_en2_4.eps}
\vskip0.5cm
\includegraphics[width=0.48\columnwidth]{files_images/Dipole_en_4.eps}
\includegraphics[width=0.48\columnwidth]{files_images/Dipole_en2_4.eps}
\vskip0.5cm
\caption{\label{fig6} Subplots (a) , (b), (c) and (d) show the same
quantities as in Fig.\ref{fig5} (a,b) but over longer
times $t>\tau$ and the x-axis is shown in linear scale. The parameters are
the same as mentioned previously in fig.\ref{fig5}. Subplot (a) shows 
the values of spring energy $U_H$ per monomer, and the bending energy $U_b$, 
per bend versus time $t/\tau$ corresponding to case-A.
 Subplot (b) shows the values of repulsive potential energy 
$U_c$, and the total energy $U_{tot}$, per monomer versus 
time  corresponding to Case A. Subplot (c) shows the values 
of spring energy $U_H$, and the bending energy $U_b$, per 
monomer versus time corresponding to Case B . Subplot (d) shows 
the values of repulsive potential energy $U_d$, and the total
energy $U_{tot}$, per monomer versus time  corresponding to
case-B.}
\end{figure}

At times beyond $t/\tau>1$, i.e. after the the helix has
already been formed, the value of  $u_H$  for the stretched
springs starts fluctuating around an average value. However,
uniform relatively uniform bends of a helical configuration are penalized by 
$u_b$  and hence at times $t>2\tau$ the uniform helical
structures start to gradually locally unwind leading to a
gradual increase in the pitch (data given later) of the 
helical structure. This can be understood by looking at the
evolution of energies $U_c, U_b$ and $U_H$  with time in
Fig.\ref{fig6}(a,b) (case A) and in Fig.\ref{fig6}(c,d)
(case B). We also show $U_{tot}$ which is 
the sum of  $U_c, U_b$ and $U_H$. There is a slight
decrease in $U_c$ or $U_d$ with time 
and the values of the bending energy $U_b$ also show a
steady but slow decrease  with time.

The next figure, Fig.\ref{fig6} shows the long time behaviour of the 
$U_c$, $U_b$ ($U_d$), $U_H$ and $U_{tot}$ as the helical conformations 
dissolve. From Fig.\ref{fig6}b and Fig.\ref{fig6}d it is evident that 
there is a crucial difference between Case A and Case B which 
arises from the difference in the rate of fall of the potential 
with increasing $r$. For Case A, $U_c$ shows a decrease of about 
$\sim 8 k_BT$ with time, whereas $U_d$ shows a decrease of about 
$\sim 2.5 k_BT$ over $20 \tau$. Consequently the total energy per 
monomer, $U_{tot}$ for the two cases also show a larger decrease 
for case A, as seen in Fig.\ref{fig6}a and Fig.\ref{fig6}c. This 
implies that the polymer chain in case A has a higher tendency to 
unwind and stretch itself out to a relatively more straight 
configuration due to  the repulsive forces of $u_c$ as compared to
polymer with potential $u_d$ of  case B. This is in spite of
the value of $\epsilon_b$, which is much higher for case B.
This  is consistent with  the data of $H2$ relaxation with time shown in Fig.\ref{fig2}(b,d) and explains why $H2$ for case B shows relatively 
higher values  as compared to $H2$ of case A at times $t > 1$.

\begin{figure}[t]
\includegraphics[width=0.48\columnwidth]{files_images/h2_h4_t_0_coulomb_dipole.eps}
\includegraphics[width=0.48\columnwidth]{files_images/h2_h4_t_0_alt_coulomb_dipole.eps}
\vskip0.5cm
\includegraphics[width=0.48\columnwidth]{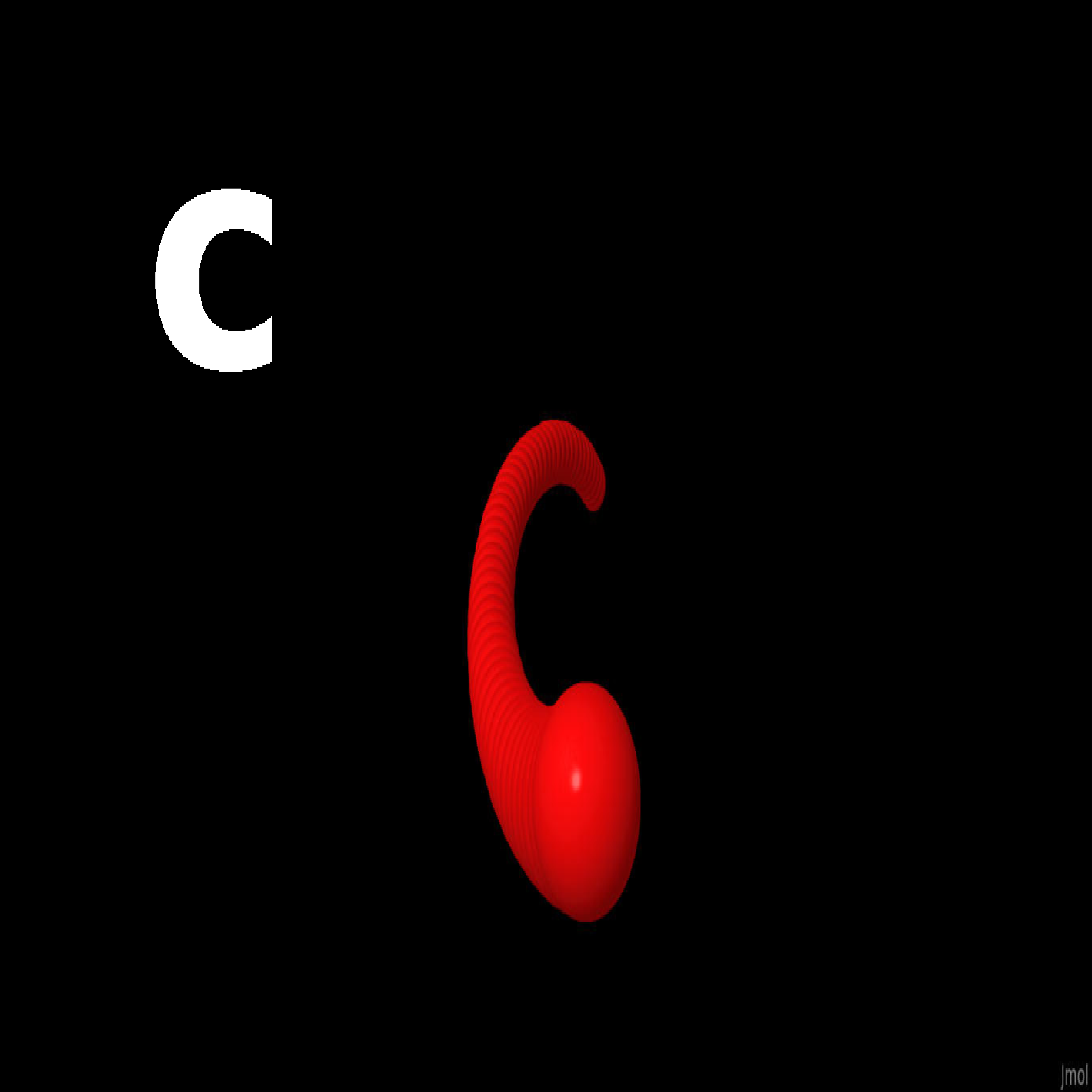}
\includegraphics[width=0.48\columnwidth]{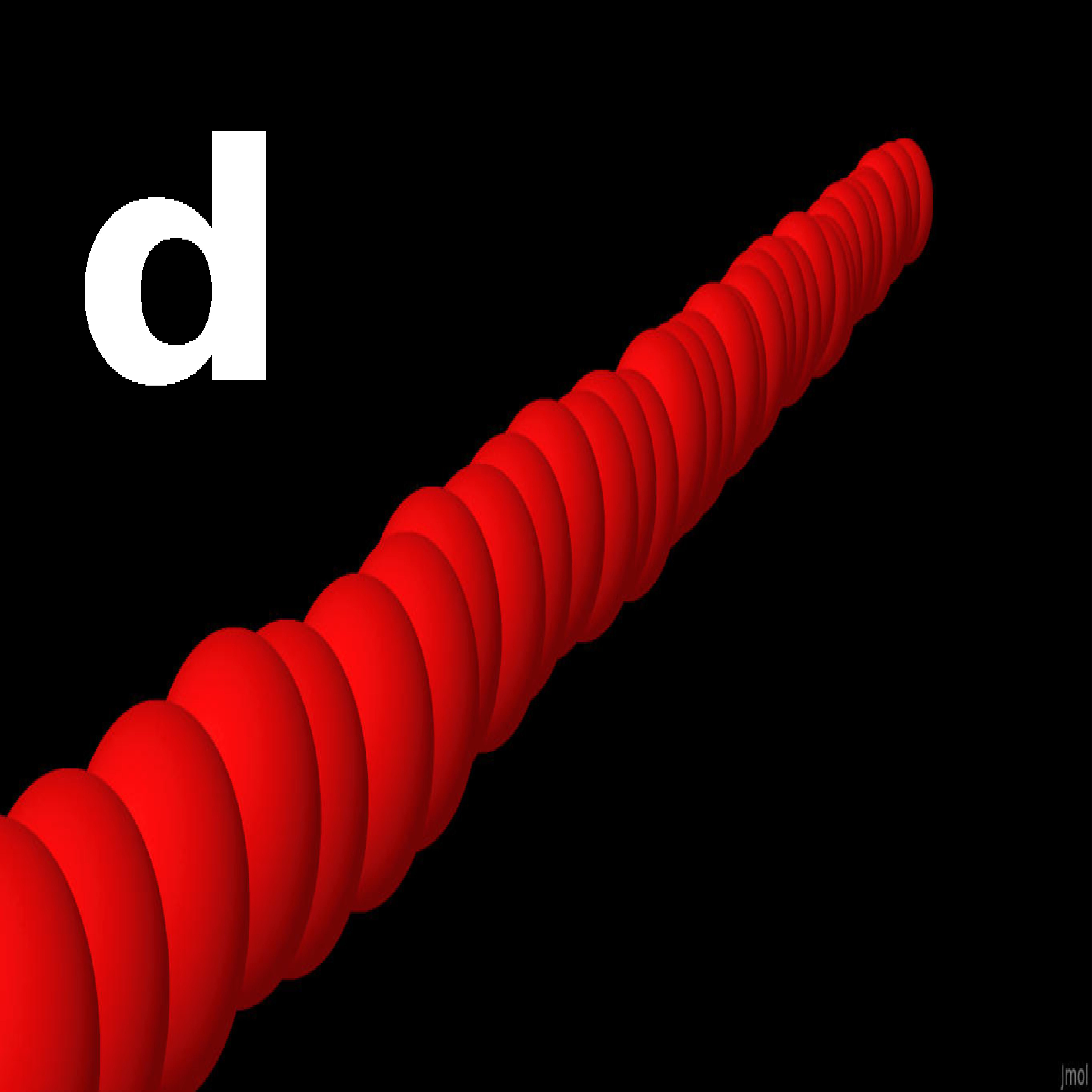}
\vskip0.5cm
\caption{\label{fig7}
 Subplot (a) shows $H4$ and $H2$ versus time a for polymer
chain corresponding to $u_c$  and $u_d$, respectively,  at 
$k_BT=0$ starting from a curved initial conformation. Subplot
(b) shows $H4$ and $H2$ versus time  for a polymer chain
corresponding to $u_c$ and $u_d$  acting between the monomer pairs, respectively  at $k_BT=0$
starting from a initial condition in which the polymer aligned
with $\hat{y}$ has small random displacements along $\hat{x}$
and $\hat{z}$.  The starting configurations for (a) and
(b) are given in subfigures (c)
 and (d), respectively. 
}
\end{figure}

\begin{figure}[t]
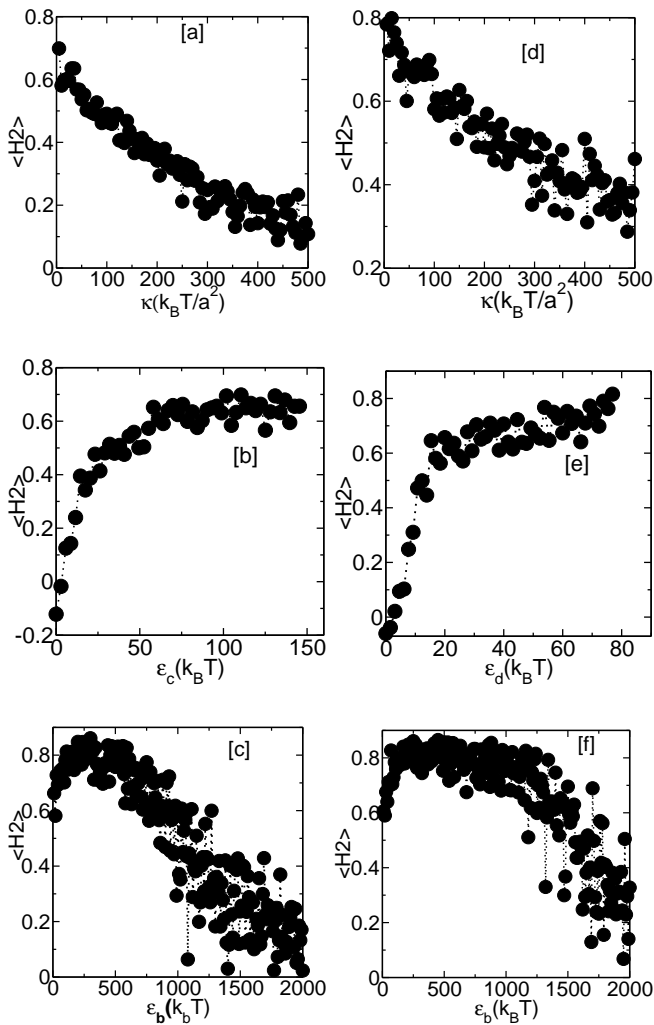

\includegraphics[width=0.49\columnwidth]{files_images/h2_kappa_phase_coulomb.eps}
\includegraphics[width=0.49\columnwidth]{files_images/h2_kappa_phase_dipole.eps}\\
\vskip0.5cm
\includegraphics[width=0.49\columnwidth]{files_images/h2_epsilon_phase_coulomb.eps}
\includegraphics[width=0.49\columnwidth]{files_images/h2_epsilon_phase_dipole.eps}\\
\vskip0.5cm
\includegraphics[width=0.49\columnwidth]{files_images/h2_bend_phase_coulomb.eps}
\includegraphics[width=0.49\columnwidth]{files_images/h2_bend_phase_dipole.eps}\\
\vskip0.5cm
\caption{\label{fig8}  Semiflexible polymer with $\kappa
=20k_BT/a^2$, $\epsilon_c=87.27k_BT$ \&  $\epsilon_b=10k_BT$
corresponding to case-A : (a),(b),(c) show change in time averaged value 
of $<H2>$ with  increase  of $\kappa$, $\epsilon_c$ $\epsilon_b$,
respectively. We change one parameter at a time keeping the other two 
parameters fixed.  Semiflexible polymer with $\kappa =10k_BT/(a)^2$, 
$\epsilon_d =107.70k_BT$ \& $\epsilon_b=80k_BT$ corresponding to
case-B:  (d),(e),(f) show change in time averaged value of $<H2>$
with increase  of $\kappa$, $\epsilon_d$ $\epsilon_b$,
respectively,  keeping two other parameters fixed.  The time 
averaged $<H2>$ was calculated over $0.66 \tau$, starting from 
$t=0.66 \tau$  to $t=1.32 \tau$.}
\end{figure}

Thermal fluctuations  provide the initial random forces
which leads to the slight displacement of the monomers away
from its initial straight line configuration which makes the
linear configuration unstable. At temperature $T=0$, the
polymer starting from a initial straight configuration along 
$\hat{y}$,  stretches out  to reach its minimum energy
configuration in presence of $U_c$ but never forms  helices 
as forces arising from $u_c$ (or $u_d$) and $u_H$ act along 
the line joining the centres of the monomers  (refer movie
S3 in Supplementary section). At temperature $T>0$ and for time 
$t/\tau >0$, the monomers move  away from the straight line 
configuration, which leads to force
components along the $\hat{x}$ and the $\hat{z}$ directions
from $u_c$ and $u_b$,  and results in the emergence of
helical conformations when $U_c \neq 0$ (or $U_d\neq 0$). To
establish these conclusions, we ran a simulations 
to calculate $H2$ and $H4$ at temperature $T=0$, however,
starting from a uniformly curved initial condition such
that the chain forms an arc in $x-y-z$ plane  (refer Fig
\ref{fig7}c).  Such an initial conformation again leads to
helical instabilities due to forces along the $x$ axis and
the $z$ axes and therefore results in helices (as seen in
data of Fig.\ref{fig7}a).  Alternatively, starting from an initial
configuration of a relatively straight polymer chain along the
$y-$ axis but with small random displacements of all monomers 
along $x$ and $z$ coordinates maintaining temperature $k_BT=0$  (refer
Fig. \ref{fig7}d), we still obtain a helical conformation of the
polymer as seen in the data of Fig.\ref{fig7}. {  Details of the initial conditions are described in 
the Supplementary section.} 

\begin{figure*}[t]
\includegraphics[width=0.99\columnwidth]{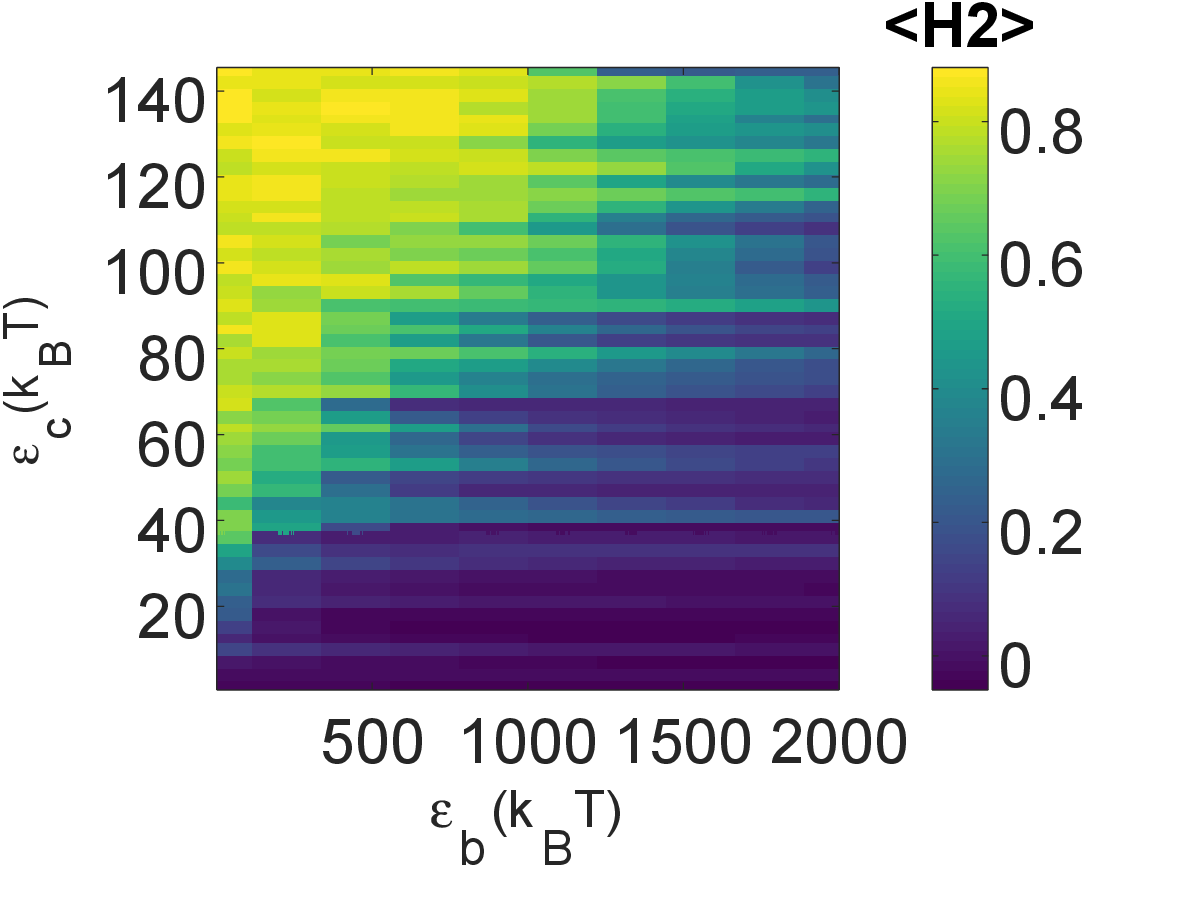}
\includegraphics[width=0.99\columnwidth]{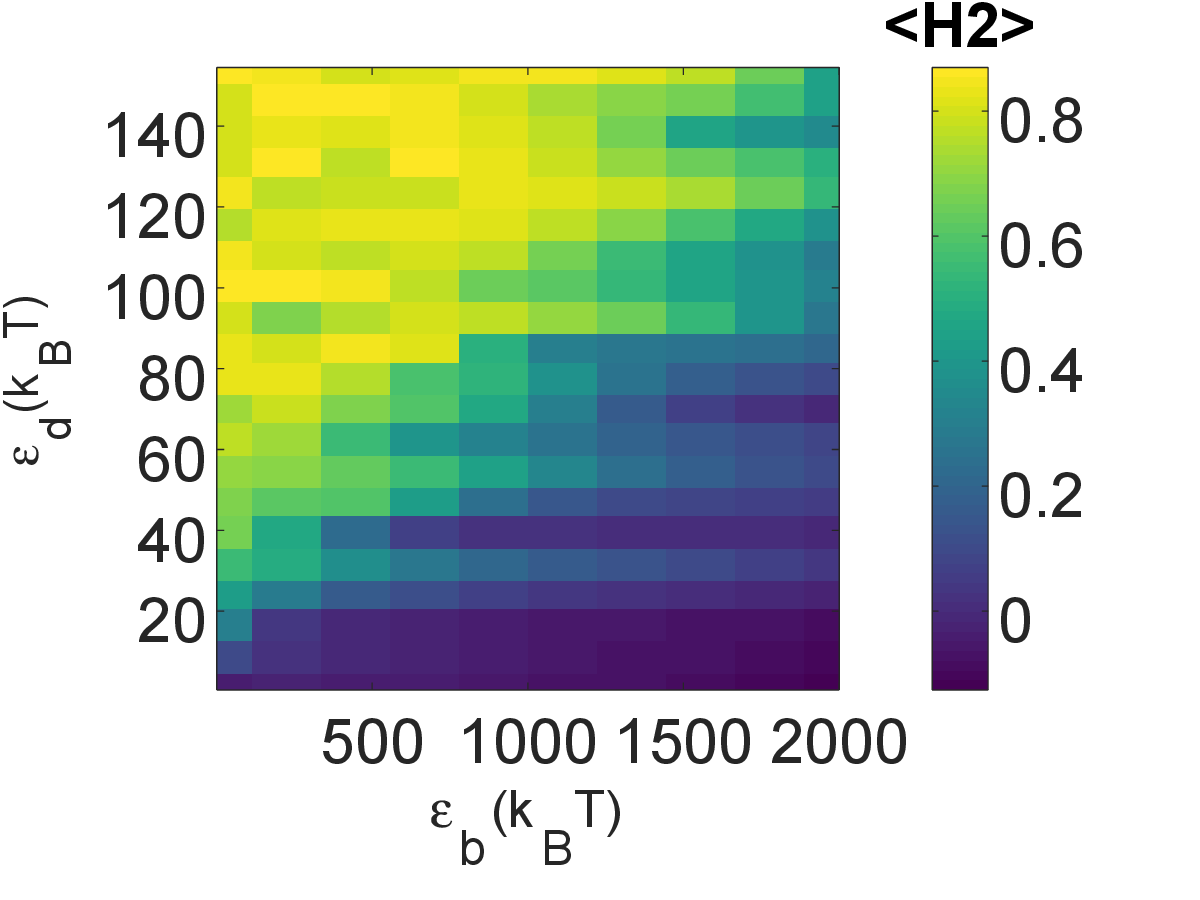}
\vskip0.5cm
\caption{\label{fig9}
Left colormap:  At $\kappa=20 k_BT/a^2$, the state diagram
shows the range of  $\epsilon_b$ and $\epsilon_c$ for which
one obtains helices. 
Right colormap: At $\kappa=10 k_BT/a^2$, the state 
diagram shows the range of $\epsilon_b$ and $\epsilon_d$ for
which one obtains helices.
The average $<H2>$ was calculated from $0.66 \tau$, to $1.32 \tau$. 
This is because we expect the helix to have been formed by $0.66 \tau$.  
When $H2<0.2$,
one can hardly distinguish between a helical polymer and a 
semiflexible polymer with bends due to thermal fluctuations.
The persistence length holds a linear relation with $\epsilon_b$ such 
$\ell_p=a \epsilon_b/k_BT$, and the charge $q$ per unit length $a$ is 
$q/a=\sqrt{4\pi\epsilon*\epsilon_c/a}$. The dielectric constant of the
medium is $\epsilon$.} 
\end{figure*}

Thus the  role of temperature is to introduce deviations
from the straight linear conformation, and this triggers the 
helical instability. Since the local helical instabilities are 
triggered by random fluctuations  at  finite $k_BT$, we do not
have any control on the handedness  of the chain at different
segments of the chain. 
Furthermore, if a single  or a couple of monomers are slightly displaced from 
a straight line configuration at $T=0$,  then  the semi-flexibility drives 
the chain to become straight  and  it then stretches out along a straight 
line to reach  its energy  minimum configuration. Thereby it does not form a  
helix { provided  the magnitude of the displacement of the 
monomers from 
the straight linear conformation of the chain is lesser than a certain value. 
To substantiate the same we ran the simulations at $k_BT=0$ for a polymer 
chain of $49$ monomers with $u_c$ acting between all monomer pairs and other
parameters pertaining to that of Case A. The simulations reveal that if 
the magnitude of the displacements from the straight linear conformation of 
the arbitrarily chosen monomers ($42nd$ and $13th$ in our case) is lesser 
than $0.0028a$ we do not obtain helices. For displacements of magnitudes 
greater or equal to that of $0.0028a$ we obtain helices. We show $H2$ and $H4$ 
versus time for a polymer chain of $49$ monomers at $k_BT=0$, with the $13th$ 
and the $42nd$ monomer displaced from the straight linear conformation by 
$0.0028a$  and other parameters pertaining to that of Case A
(refer Supplementary).

For a different choice of displaced monomers, the minimum displacement essential for 
helix formation will change. This is because the monomers of the polymer chain
experience different net repulsive forces depending on their relative positions 
with respect to other monomers. It is also to be noted that for higher values 
of semi-flexibility a larger magnitude of the displacement of the monomers 
from the straight linear conformation would be required for the repulsive 
interactions to overcome semi-flexibility}. We emphasize that  the helical
configuration at $T>0$  is not a energy minimum state but a configuration 
that the polymer accesses in its kinetic pathway to its (free)
energy minimum state which is  a stretched straight configuration(s) of monomers with  local bends depending  on the relative strengths of $u_b$ and $k_BT$.


At non-zero $k_BT$, if the soft spring ($\kappa=10k_BT/a^2$ and
$\kappa=20k_BT/a^2$ for cases A and B, respectively)  joining the
monomers becomes too stiff then the  the position of monomers do not 
time-evolve to form a helix  in response   to forces arising from  
$u_c$ and $u_d$. For high values of $\kappa$, stiff springs do not permit  
monomers to radially stretch out locally, thus preventing helix
formation. We refer the reader to  Figs.\ref{fig8} (a,d)
which shows the decreasing values of the $<H2>$ order
parameter  with increasing values of $\kappa$. The angular 
brackets in $<H2>$ denote the average value of $H2$ calculated using data 
collected  between $0.66\tau$ to $1.32 \tau$.
On the other hand, increase in the value  of $\epsilon_c$ 
in $u_c$ (or $\epsilon_d$ in $u_d$) increases  propensity 
of helix formation as observed in the increase in the value
of $<H2>$ with $\epsilon_c$ (or $\epsilon_d$) in
Figs.\ref{fig8}(b,e). For values of $\epsilon_c>50 k_BT$ and
$\epsilon_d>20 k_BT$, $<H2>$ nearly saturates to values of
$~0.75$. There is no helix formation when $u_c, u_d=0$.

High values of $\epsilon_b$ in the expression for $u_b$ 
hinder the formation  of sharp kinks  which subsequently
stretch out radially to give rise to the helical 
structures, thereby, suppresses the instability: refer 
Figs.\ref{fig8}(c,f) corresponding to cases with potentials
$u_c, u_d$.  We get finite values of $H2$ even when
$\epsilon_b =0$, as a charged polymer chain with the same 
sign of charge on the monomers behaves like a semi-flexible chain \cite{Manning2006,Trizac2016,Dobrynin2005,Li1995} .
Hence, an increase in the values of $\epsilon_b$ from zero 
leads to an initial increase of $<H2>$ as increased bending energy costs
help in radially spreading out the polymer as it leads to reduction of
bending energy. Thereby, $<H2>$ reaches a peak value of $0.9$ at intermediate
$\epsilon_b$ values. But thereafter, $<H2>$ starts decreasing
with further increase of $\epsilon_b$ as reasoned earlier.

Thus only in a certain range of these interactions of $u_c$
(or $u_d$) and $u_b$ do we obtain well formed helices.
This is further illustrated by the two state diagrams  shown in
Figs.\ref{fig9} which map out the average values of $H2$
for various combinations of the values of $\epsilon_c$ (or
$\epsilon_d$) and $\epsilon_b$. To obtain  the colormaps
shown in Figs.\ref{fig9}(left) and \ref{fig9}(right),
$\kappa$ was fixed at the same values as given previously
corresponding to case A and case B. The colormaps in
Fig.\ref{fig9}- indicate that for higher values of
$\epsilon_c$ (or $\epsilon_d$), helices can be obtained for relatively higher
values of $\epsilon_b$ because the helix formation depends on 
the relative strengths of $u_c$ (or $u_d$) and $u_b$.

A polymer with relatively very high values of $\epsilon_b$
is unable to form helices as formation of sharp kinks will 
be prevented by  very high bending energies.  Kinks, even if 
formed , will relax to form configurations which are 
stretched out resulting in lower values of $H2$ and higher 
values of pitch (as discussed later). The colour map in the 
subfigure on the right of Fig.\ref{fig9} also shows that case B
leads to formation of helices even with higher values of
$\epsilon_b$ as compared to that in Case A. This is because,
the polymer with $u_c$ (case A) would cause the polymer contour
to stretch out more with relatively larger pitch during helix
formation at times $t < \tau$ due to longer range of $1/r$ potential 
as compared to that of $1/r^3$. This results in lower  values of $H2$
when $\epsilon_c$ is high as compared to a polymer with $u_d$
potential with similar high  values of $\epsilon_d$. Similar arguments
were discussed previously when discussing the long time
relaxation of the helices in Figs.\ref{fig6}(b,d). 
So when comparing helix formation in case A with case B 
with relatively large values of $\epsilon_c, \epsilon_d$ (say, with the choice
$\epsilon_c = \epsilon_d$), $H2$ values will be lower in case A, as helix 
formation will be suppressed more in case A than in case B at identical 
high values of $\epsilon_b$.

We have already discussed how $\epsilon_b$ is related to the persistence length.
Similarly, we must express $\epsilon_c$ in terms of line
charge density of a polymer. We remind the reader that 
for case-A the Coulomb repulsion between two similarly charged monomers of charge $q$ placed at a distance of $a$ from each
other, is equal to $\epsilon_c$.
Therefore,$(q^2/4\pi\epsilon a)=\epsilon_c$, where $\epsilon$
denotes the dielectric constant. Hence, 
\begin{equation}
q/a=\sqrt{4\pi\epsilon*\epsilon_c/a}.
\end{equation}
As a reference, if the distance between monomers is $a=10 nm$,
then at $T=300K$, $\epsilon_c=87.27k_BT$  corresponds to a charge
of  $\approx 36e$ on each monomer $10 nm$ apart in water. This corresponds to a polymer
chain having a charge density lower than the charge density of DNA.
Bare DNA has around $22 e$ charge in a $3.7$ nm segment \cite{milo2015cell}. So such transient helical configurations can be seen in DNA or 
polymers with line-charge densities lower than that of the DNA, if they become charged from a neutral configuration.

\begin{figure}[t]
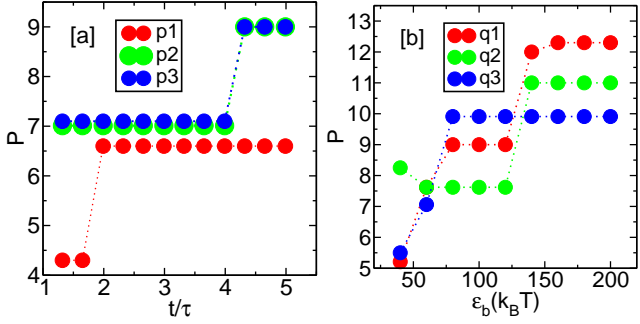

\includegraphics[width=0.48\columnwidth]{files_images/pitch_vs_t_npart_100_coulomb.eps}
\includegraphics[width=0.48\columnwidth]{files_images/pitch_vs_bend_npart_100_coulomb.eps}
\vskip0.5cm
\caption{\label{fig10}
(a) Plot of  the (most freqently occurring) pitch $P$ versus time for a chain of $100$ monomers
for $\kappa=20k_BT/a^2$, $\epsilon_b=10k_BT$ \& $\epsilon_c=87.27k_BT$ (Case A) 
for $3$ independent  runs indicated by $p1$, $p2$ and $p3$. (b) At
a fixed time $t=\tau$, we plot the (most frequently occurring) pitch $P$ versus  $\epsilon_b$
for $3$ independent runs $q1$, $q2$ and $q3$. The other parameters correspond to that of Case A. The figures corresponding to Case B is given in the supplementary. 
}
\end{figure}

What determines the pitch of the helix and how can we control 
it? The procedure for calculating  the most frequently occurring pitch (in units of monomers) 
is detailed in the  supplementary section.  Once
formed, the pitch of the helix  increases with time as the
helical structure gradually unwinds over time to decrease  bending 
energy costs. To that end, we show the variation of the (most frequently occurring)  pitch $P$ versus 
time in Fig.\ref{fig10}a for a chain length of $N=100$ monomers for 
a semiflexible polymer with $u_c$ acting between the monomers.
{A 
polymer with large $N$ is chosen so that the fourier transform 
calculation yields more accurate results (refer supplementary).} 
The quantity $P$  increases with increasing  $\epsilon_b$ (refer 
Fig.\ref{fig10}b) as a higher value of $\epsilon_b$ results in
a higher energy cost associated with the local bends along the 
polymer chain. Thus a higher value of $\epsilon_b$ gives rise to
fewer loops along the chain, or a higher average value of the pitch. We did not observe any significant dependence of the pitch on $\kappa$ and 
$\epsilon_c$ (or $\epsilon_d$). The corresponding data for $P$ (versus time and $\epsilon_b$) with $u_d$ acting between all monomer pairs (Case B)
are given in the supplementary. 

 In Fig.\ref{fig10}a we see that the quantity $P$  is constant over 
 some time before it abruptly shifts to a higher value. We explain how $P$ is calculated to understand why that is the case. 
 When we calculate the pitch,
 we take  the Fourier transform of the quantity $W$, which is the dot product of bond vectors along the contour with a vector perpendicular to the axis of the helical polymer chain. Refer supplementary
 for the procedure of calculating $P$ and also  refer the figures 
 which show the different values of the pitches obtained in the same helical configuration as it evolves with time. 
 { Thus, different segments of the chain form helical 
 structures with  slightly different values of the pitch. These 
 in turn unwind at different rates.   Hence there is more than one peak 
 in the Fourier Spectrum of `W' at any given instant of time. 
 The monomer index corresponding to the
 peak with the highest amplitude at any given instant of time is denoted 
 as $P$. Thus  $P$  represents the most frequently occurring pitch in 
 the helical polymer chain. 
 As the helical structure gradually unwinds segment by segment, 
 the pitch  corresponding to a particular segment increases and consequently the
 amplitude of the corresponding peak in the Fourier spectrum changes.
 However there is a change in the quantity $P$ for the entire polymer 
 chain, only when the position of the highest peak in the Fourier Spectrum of 'W changes. This is evident from Fig.6b of the Supplementary where the Fourier 
 spectrum  shows two significant peaks at time $t/\tau=1$. The amplitude of 
 the peak corresponding to a pitch of $9$ monomers, gradually increases with 
 time, until at time $t/\tau=4.33$, the peak corresponding to $9$ monomers 
 becomes the peak with the highest amplitude. It is only at this point that 
 we register a change  in the value of $P$ of the helical polymer chain. 
 Thus,the most frequently occurring pitch in the helical polymer chain or $P$ therefore shows abrupt jumps in time.}

 In Fig.\ref{fig10}b, where we show the dependence of $P$  with
 $\epsilon_b$ we choose not to calculate the ensemble mean, since
 there are large differences in the values of  $P$ at a fixed
 instant of time 
 corresponding to different runs.
 To illustrate this point we show three independent runs, which
 show considerable differences in the value of pitch, at the same
 time and for the same value of $\epsilon_b$. The difference in
 the values of $P$ for independent runs arise due to the fact that
 initially the helix formation is triggered by the presence of
 thermal fluctuations.

 As we saw earlier, that the formation of the helix depends 
 on the strength of the Coulomb interaction $\epsilon_c$ or
 on the  value of $\epsilon_d$.  The question is if the value of 
 $\epsilon_c$ in the model  polymer chain  increases gradually with time, 
 i.e. a neutral  semi-flexible chain gradually becomes charged 
 (e.g. { say}, 
due to change  in pH), { does the polymer still form a helix 
if it starts out from a  relatively straight configuration 
in the presence of thermal fluctuations?} Moreover,  can the helix formation 
occur in a recurring manner as a response to a time dependent periodic 
repulsive interaction?

\begin{figure}[t]
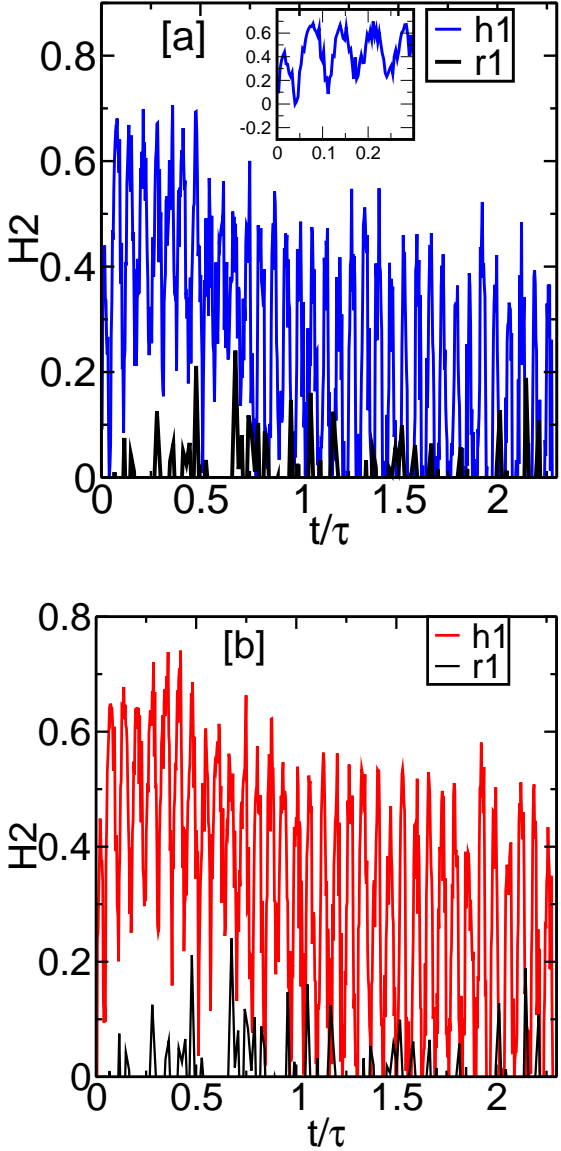

\vskip0.2cm
\includegraphics[width=0.85\columnwidth]{files_images/h2_driven_coulomb.eps}
\vskip0.7cm
\includegraphics[width=0.85\columnwidth]{files_images/h2_driven.eps}
\vskip0.5cm
\caption{\label{fig11}
 Subfigure (a) shows  $H2$ versus time $t$, scaled by the
 relaxation time $\tau$, for a chain of $49$ monomers with a time
 dependent $U_c(t)= \epsilon_c^0*(a/r)\cos^2(2 \pi t/T_0)^2 $ with
 $T_0=0.13\tau$ over many cycles. $H2$ varies periodically nearly
 in phase with the forcing. The inset shows that the periodicity
 is $T_0/2$. (b) Subfigure shows $H2$ versus time for   $u_d(t)$
 which has a similar time dependence as $u_c(t)$.
 We also give $H2$ data for when $u_c=0$ and $u_d=0$, denoted in 
 the legend as $r1$  for 
 comparison of response.}
\end{figure}

To that end, we choose a significantly more  rigid  polymer  such 
that the persistence length is much larger than the contour length of the polymer chain  with $N=49$ monomers. We also choose a  suitably higher 
charge density of the polymer chain. Moreover, we use a time dependent 
potential of the form $u_c(t)= \epsilon_c(t)(a/r)$ where 
$\epsilon_c(t) = \epsilon_c^0*\cos^2(2 \pi t/T_0)$ where we have 
chosen $T_0=0.13\tau$, $\epsilon_c^0=727.3k_BT$ and $t$ denotes the
simulation time. The values of $\epsilon_b$  and $\kappa$  was
changed to $300k_BT$ ($\ell_p=300a$) \& $1500 k_BT/a^2$, respectively, 
to have a stiffer chain. We observe that we obtain helices, in a
recurring manner. The helices form, then dissolve away as 
$\epsilon_c(t)$ becomes  zero, such that  the polymer becomes 
relatively straight in the thermal bath. The helical conformation forms
again as the amplitude of the periodic forcing increases. 
To substantiate that we show $H2$  versus time $t$ in Fig.\ref{fig11}a
for a chain of $49$  monomers under the influence of $u_c(t)$ and 
also for a chain of $49$ monomers  such that $u_c=0$. We also show 
data for the same values of $\kappa$ and $\epsilon_b$ but $u_d (t) =
769.34(a/r)^3*\cos^2 (2 \pi t/T_0)$ in Fig.\ref{fig11}b, where again 
we obtain   helices in a recurring manner.  The helices formed for 
these high values of $\kappa$ and $\epsilon_b$ dissolve away 
significantly faster as compared to that of cases A and B, and 
quickly return to  a relatively straight conformation.
{ This  again emphasizes that the helix formation 
does not critically depend upon the straight line initial condition provided
$\ell_p$ is  larger than the contour length; a factor of 6 in this case.  
Each time the polymer straightens up before reforming the helix, the 
configuration is slightly different due to the presence of $k_BT$.}

Thus for the value of $T_0$ chosen   for this study, the helices can 
be made to form and dissolve away in  a recurring, periodic fashion.  
{ The time  scale $\tau$ is decided by  the value of the friction 
constant  $\zeta$.} Finally we want to put the relatively high value of 
$\epsilon_c$ in  perspective. If the distance between  monomers in our 
calculation is taken  as $a=10 nm$, then at   $T=300K$, the Coulomb energy 
$\epsilon_c=727k_BT$  used for the above study  corresponds to a 
charge of  $\approx 108 e$  on each monomer in water, i.e. a line charge 
density of $\approx 11e$  per nm. As a reference, each base pair  of DNA of size
$\approx 3.4$\AA \,  has a charge  of $\approx-2e$ at physiological 
pH \cite{milo2015cell}.  Thus our choice of $\epsilon_c$ in this case  leads to a 
line charge density twice than that of DNA.

\begin{figure}[!hbt]
\includegraphics[width=0.48\columnwidth]{files_images/coulomb_ends_fixed_h2.eps}
\hskip0.1cm
\includegraphics[width=0.48\columnwidth]{files_images/coulomb_ends_fixed_h4.eps}
\vskip0.5cm
\includegraphics[width=0.48\columnwidth]{files_images/dipole_ends_fixed_h2.eps}
\hskip0.1cm
\includegraphics[width=0.48\columnwidth]{files_images/dipole_ends_fixed_h4.eps}
\hskip0.1cm
\includegraphics[width=0.68\columnwidth]{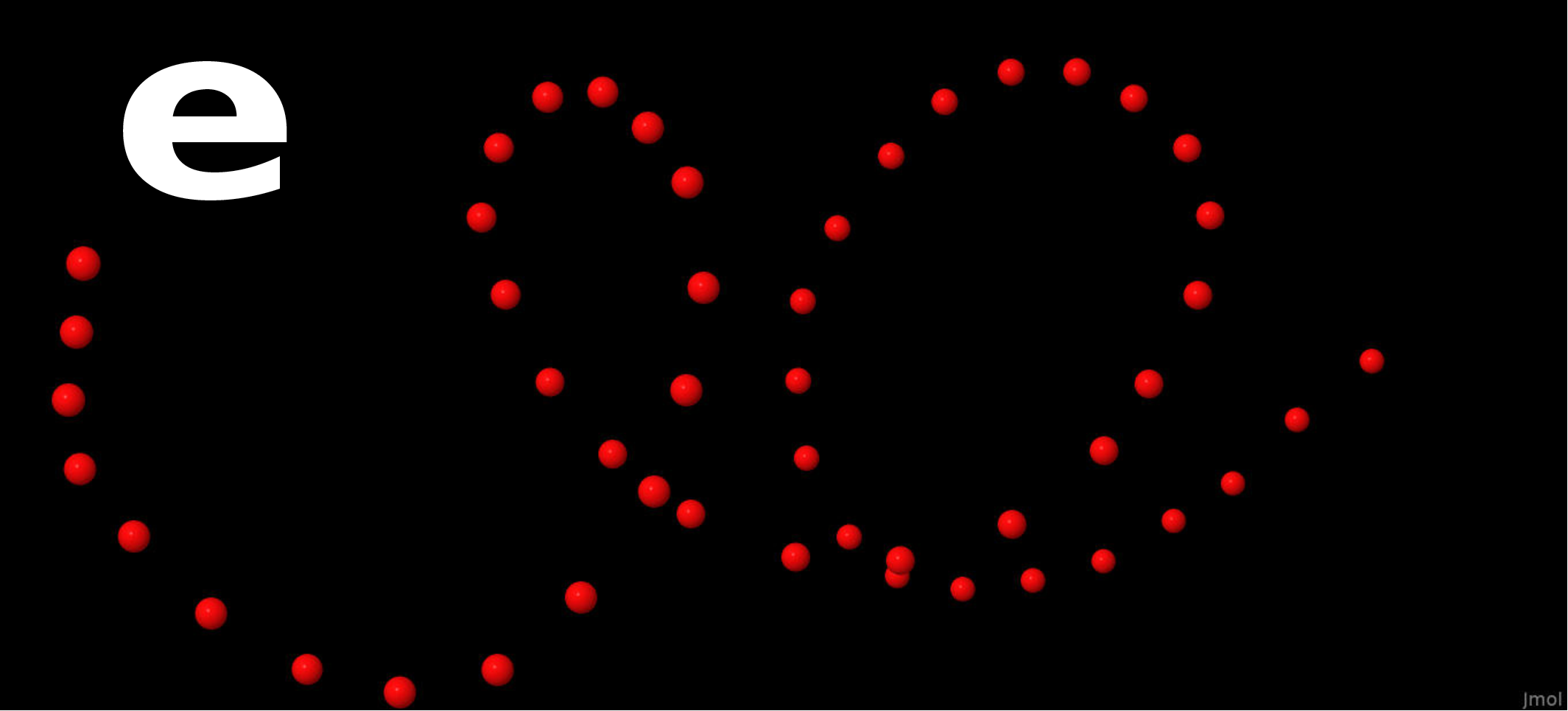}
\hskip0.1cm
\caption{\label{fig12}
{ Subplots (a) and (b) shows the evolution of order parameter 
$H2$ and $H4$ 
versus time for a semi-flexible polymer chain of $49$ monomers with its end monomers fixed and forces due to  potential $u_c$ acting between all monomer pairs,
with all parameters being identical to that of Case A. Subplots (c) and (d)
shows the evolution of o $H2$ and $H4$  versus time for a semiflexible polymer
chain of $49$ monomers with its end monomers fixed and forces due to potential $u_d$
acting between all monomer pairs, with all parameters being identical to that of
Case B. Subplot (e) shows the snapshot of a polymer chain of $49$ monomers at
$t=20\tau$, with end monomers fixed and $u_d$ acting between all monomer pairs and other parameters
pertaining to that of Case B  ($\mathbf{H2=0.84,H4=0.50}$)}.
}
\end{figure}

{ 
To explore whether the helical structure becomes more long-lived when the two 
ends of the polymer chain are grafted (tethered) on to two parallel surfaces, 
we do not update the positions of the end monomers of a polymer chain while 
observing the dynamics of  the chain.
The distance between the fixed monomers is equal to the contour length of the polymer chain (of $49$ monomers) in the absence of charges. In this case the helical
structures persist for a longer duration of time as compared to the helical
structures resulting from a free standing polymer. This can be surmised 
from the data given in  Fig.\ref{fig12}. With $u_c$ acting
between the monomers, Figure \ref{fig12}a shows that there is a slight 
increase in the value of $H2$ for a polymer chain  at longer times (e.g. at time 
$t=20\tau$)  as compared to a free standing polymer chain at similar times
refer Fig.\ref{fig2}b. Moreover, Fig.\ref{fig12}c  shows that there is a 
significant increase in the value of $H2$ at long times (at $20\tau$)  as
compared to an free standing polymer chain with $u_d$ acting 
between the monomer pairs at similar long times, refer Fig.\ref{fig2}d. 
Thus we conclude that the tethering hinders the relaxation of transient helical
structure by preventing it to stretch axially. 
The effects are more pronounced with the  interaction potential $u_d$. 
A snapshot of the helical conformation of a polymer chain of $49$ monomers  with end monomers fixed  and $u_d$ acting between the monomer pairs has also 
been provided in Fig.\ref{fig12}e.} 

\begin{figure}[!hbt]
\includegraphics[width=0.8\columnwidth]{files_images/lj.eps}  
\includegraphics[width=0.8\columnwidth]{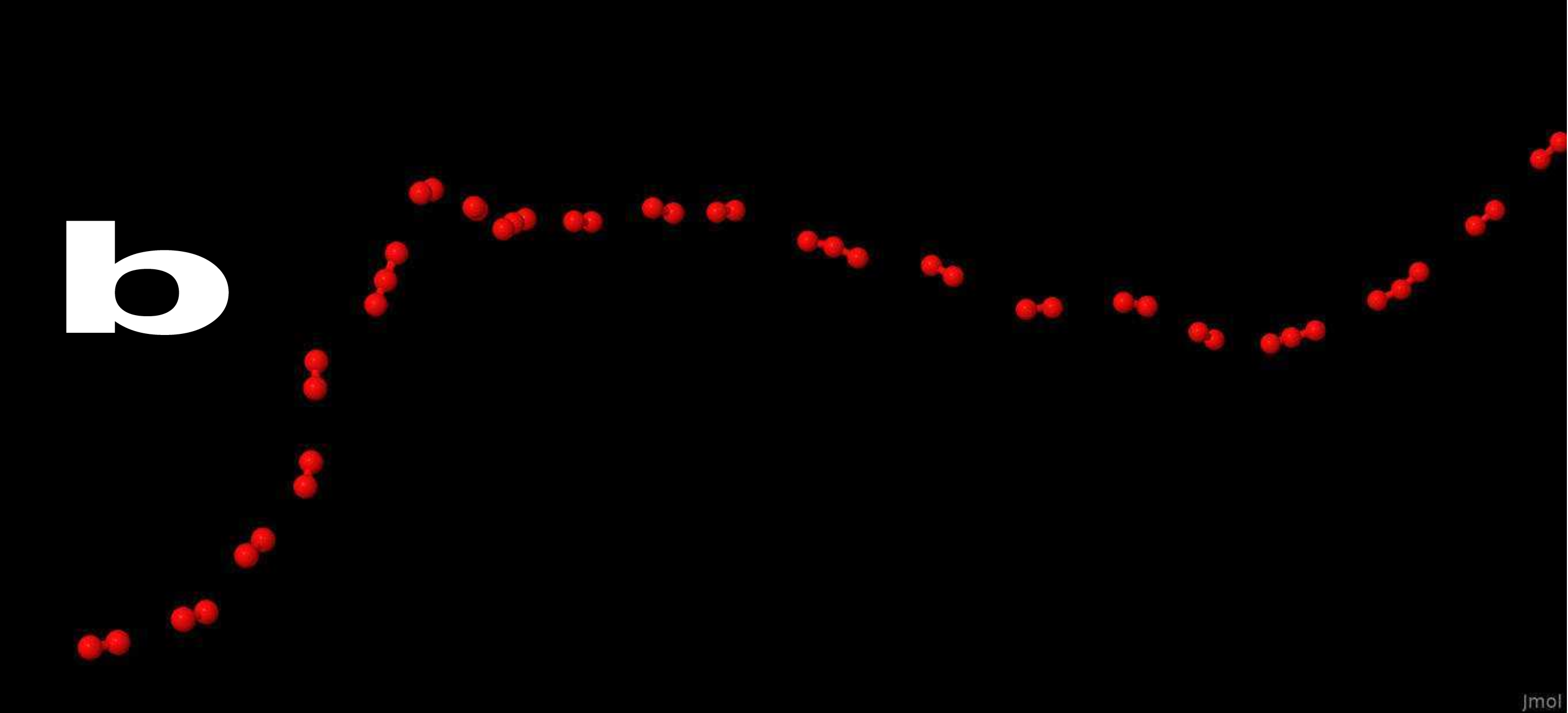}  
\caption{\label{fig13}
   Subplot (a) shows $H2$ versus time for two different values of $k_{int}=\epsilon_{lj}/\epsilon_c$ (where $\epsilon_{lj}/4$ is the depth of the Lennard Jones potential) with three independent runs corresponding to each value. All parameters are identical to  that of Case A. 's1','s2' and 's3' denote three independent runs for $k_{int}=2.29$, while 'w1', 'w2' and 'w3' denote three independent runs for $k_{int}=3.43$. Subplot (b) shows the snapshot of the configuration of the polymer chain of $49$ monomers for $k_{int}=3.43$ at $t=230\tau$.  
 }
\end{figure}

{ In our simulations so far we have implicitly assumed the
solvent to be  a good solvent. To investigate if the solvent quality 
affects helix formation,  we present data for simulations with polymer 
in bad solvent conditions. To model bad solvent conditions, 
we apply an attractive Lennard Jones (LJ) interaction (of  potential depth
$=\epsilon_{lj}/4$). This is used  in conjunction  with the repulsive Coulomb interaction  $u_c$ with all parameters pertaining to that of Case A
to study transient helix formation. A polymer in a bad
solvent  would lead to a collapse 
of the  polymer, where as the Coulomb repulsion would keep the polymer in 
a stretched  condition. We show that as long as strength of attractive
interaction is relatively low as compared to the repulsive Coulomb interaction, we manage to obtain helices. 
If the ratio of the Lennard Jones interaction strength to the
strength of the repulsive interaction i.e ($k_{int}=\epsilon_{lj}/\epsilon_c$), 
is greater than a certain critical value, then the helix formation is 
prevented.  For a polymer chain of $49$ monomers with $u_c$ ($\epsilon_c=87.27 k_BT$) acting between all monomer pairs 
and other parameters kept identical to that of Case A, if $k_{int}$ is lesser 
than $k_{int}=3.43$, only  then do we obtain helices.
To substantiate the same, 
we have Fig.\ref{fig13}  where we show $H2$ values versus time for a 
polymer chain of $49$ monomers  with $k_{int} =2.29$ ( `s1',`s2' and `s3'
correspond to independent runs)  while `w1', `w2' and `w3' denote three
independent runs with $k_{int}=3.43$. We note that for all the three 
runs the value of $H2$ is significantly greater for $k_{int}=2.29$. 
For $k_{int}=3.43$ one obtains a long lived configuration with small 
clusters of monomers separated by stretched springs as shown in Fig.\ref{fig13}b. 
A detailed study of the effect of unscreened Coulomb interaction and 
polymer collapse due to bad solvent conditions, and how the minimum value required for helix formation, $k_{int}^c$,
depends on the chain length can be explored in a future study.}

\section{Discussions and Outlook} In conclusion, we demonstrate
that spherically symmetric long ranged {\em repulsion} can 
give rise  to { transient} helices in a semi-flexible polymer 
This is a 
consequence of the long range of the interactions used which helps
to radially spread out the sharp kinks that are formed at short times by the polymer chain 
due to a  combination of  thermal forces  and repulsive
interactions between monomers.  Importantly, we have considered 
the charges on the polymer chains to be
unscreened by counterions.  Our model is minimal by design 
and therefore  doesn't take into account atomistic chemical 
details of the monomers or the solvent particles. { We find our findings
non-intuitive apriori, because in previous studies emergent helices (in 
the absence of torsional potentials) are observed typically as a consequence 
of packing effects due to confinement or energy minimization due to short ranged
attractive interactions in filaments, where sharp kinks are explicitly prevented.} 

The transient helix formation { that we observe} cannot be analyzed 
using geometric  or a energy minimization calculation  
as the minimum (free) energy
configuration   in the presence of Coulomb potential $u_c$ (or $u_d$)  
is not a helix;  it is  a straight line configuration with
deviations  due to   thermal fluctuations. However, a uncharged polymeric chain,
which is slightly  perturbed from a straight line  initial condition 
or  is in a bent configuration at $T=0$, is put in conditions such that 
the charge on the 
monomer gets switched on at a time $t=0$, it  relaxes to equilibrium
through a kinetically driven  pathway where the intermediate 
stage is a helical configuration. This observation remains true even 
if we start out with a stiff polymer in thermal equilibrium with a 
solvent bath. The same phenomenon happens 
even if the monomer charge  increases gradually from zero as 
shown in Fig.\ref{fig11}.  Interestingly,  we can get the helix  
to form in a recursive fashion as has been 
 demonstrated in  Fig.\ref{fig11}   as the charge is gradually
 increased and then decreased back to zero in a periodic manner.

 { Since a free standing charged polymer chain tends 
 to stretch out axially at long times, we can also use the charging 
 and  discharging of a (tethered) polymer chain to apply forces at the 
 two surfaces to which the end monomers are kept attached.}
 We  obtain transient helices also on switching on a repulsive $1/r^3$
 potentials  for a stiff polymer in a thermal bath as long as the 
 persistence length is greater than the contour length  of the polymer chain.
 { Since the charge densities required to see the transient 
 formation is very much realizable in the laboratory, we hope that our study 
 will spur future experiments}.

Our proposed  mechanism can be  possibly used to design
helical springs for NEMs/MEMs devices at $10nm-\mu$ length scales and using
material of choice by arresting the relaxation process at a
suitable time. { As an example, we have shown that we obtain
relatively long lived-helices by fixing both the ends of 
the chain and switching on the repulsive interactions between the monomers. 
In this case the helical structures persist for a longer duration of time 
as compared to the helical structures resulting from a free standing polymer,
especially when we use $1/r^3$ interaction potentials.}
{ Since the relaxation  time of the polymer chain depends on the 
friction constant $\zeta$, a charged polymer can be made to relax slowly 
by placing it in a  solvent of higher viscosity.} { }

\vskip0.5cm
We thank K.Guruswamy and Bipul Biswas for useful discussions. We have used  
computer cluster obtained using DBT Grant BT/PR16542/BID/7/654/2016 to AC. AC
acknowledges funding by DST Nanomission, India, the Thematic Unit Program (Grant
No. SR/NM/TP-13/2016), MTR/2019/000078 and discussions in Stat-phys meetings in 
ICTS, Bangalore, India.

\appendix
\section{Persistence length}
\begin{figure}[hbt!]
\includegraphics[width=0.3\columnwidth,angle=0]{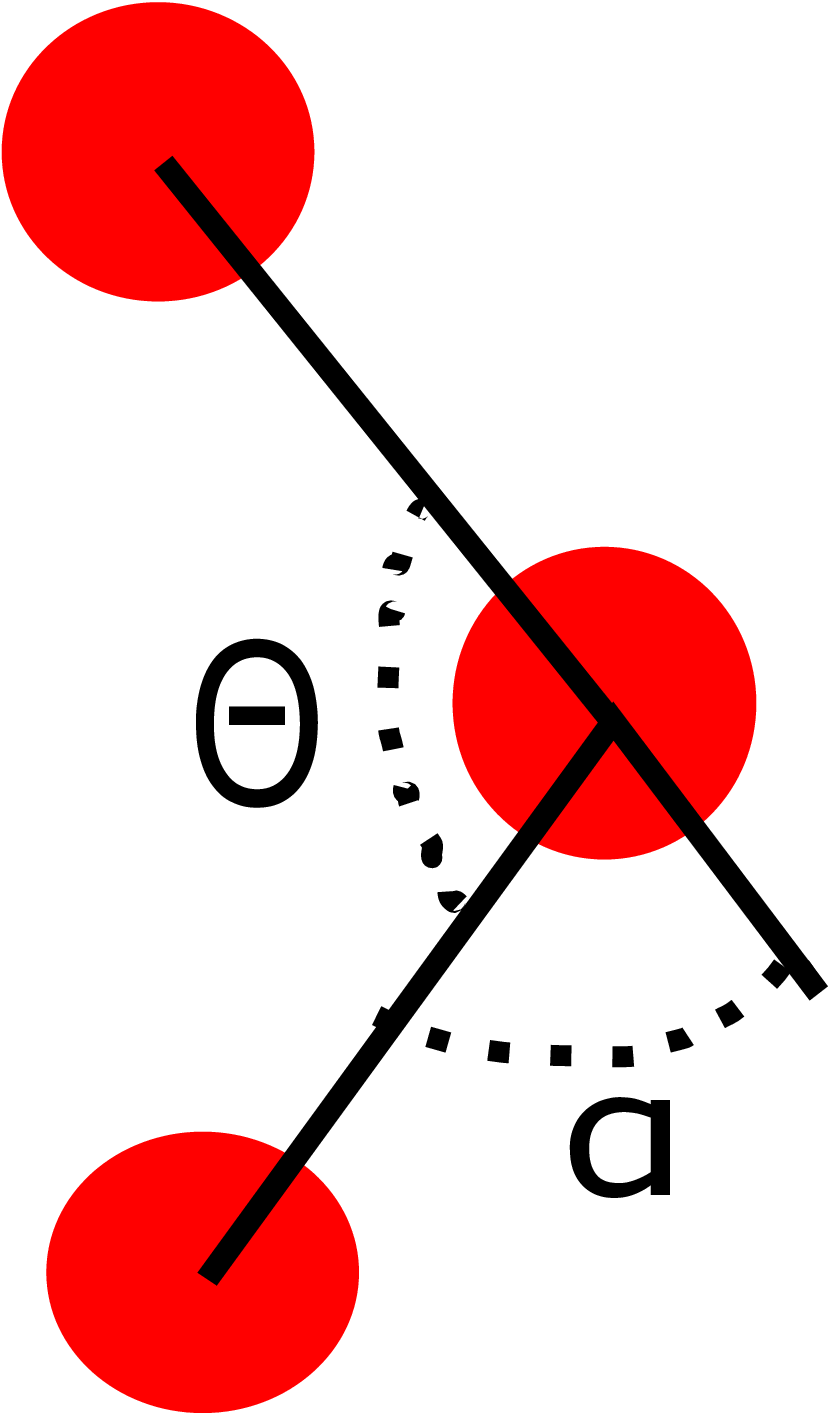}
\caption{\label{fig14} A schematic diagram showing a triplet of monomers in red 
and defining the angle $\theta$ and $\alpha$ for 
the convenience of the reader.}
\end{figure}
\vskip0.5cm
If we have a semi-flexible polymer chain with just the harmonic spring interaction 
$u_H$ and the potential $u_b=\epsilon_b \cos \theta$ which  introduces 
semi-flexibility along the chain contour then, the energy required to bend a 
triplet of monomers of semi-flexible chain from its straight line configuration (such
that $\theta_0=\pi$ and energy $u_b=-\epsilon_b$) to a configuration with 
$\theta < \pi$ is provided by the thermal energy. Therefore, if we equate the 
bending energy with the thermal energy and choose  $k_BT =1$ as we use $k_BT$ as 
the unit of energy:

\begin{equation}
\epsilon_b (cos(\theta)-cos(\pi) )\approx k_BT 
\end{equation}

\begin{equation}
 \equiv  \cos \pi - \cos \theta =\frac{-1}{\epsilon_b'}.
\end{equation}
where,
\begin{equation}
 \epsilon_b'={\epsilon_b/k_BT}.
\end{equation}
If we define $\alpha= (\pi -\theta)$, then 
\begin{equation}
-\cos \theta=\cos \alpha=(\epsilon_b'-1)/\epsilon_b'
\end{equation}
For small values of $\eta$, one can write:
\begin{equation}
(1 - \frac{\alpha^2}{2}) = (\epsilon_b' -1)/\epsilon_b'
\equiv  \alpha^2 = 2/\epsilon_b'
\end{equation}
From polymer physics \cite{Rubinstein},
we know that for  WLC (worm like chain) model, for the small 
angles  of bends, the persistence length $\ell_p$ is given by
$\ell_p = 2a/\alpha^2$
Then using Eqn.3, the persistence length 
\begin{equation}
\ell_p = a \epsilon_b/k_BT.
\end{equation}

\bibliographystyle{unsrt}
\bibliography{ref.bib}
\end{document}